\documentclass[twocolumn,times,tighten]{aastex63}
\usepackage[utf8]{inputenc}
\usepackage{amsmath,amstext,amssymb}
\usepackage{rotating}
\usepackage{booktabs}
\usepackage{numprint}
\usepackage{graphics}
\usepackage{graphicx, float}
\usepackage{epsfig}
\usepackage{wrapfig}
\usepackage{xcolor}
\usepackage[figure,figure*]{hypcap}
\usepackage{natbib}

\definecolor{Green}{rgb}{0,0.5,0}
\definecolor{Blue}{rgb}{0,0,1}
\definecolor{Red}{rgb}{1,0,0}

\newcommand{\kms}{km s$^{-1}$}
\newcommand{\etal}{$et al.$} 

\newcommand{\msun}{M$_{\odot}$}
\newcommand{\degrees}{\ensuremath{^\circ}}
\newcommand{\degree}{\degrees}

\newcommand{\aone}{ALFALFA}

\def\kms{km~s$^{-1}$}

\received{July 30, 2020}

\submitjournal{AJ}

\shorttitle{ALFALFA-SDSS}

\shortauthors{Durbala $\etal$}

\begin{document}

\title{The ALFALFA-SDSS Galaxy Catalog }

\correspondingauthor{Adriana Durbala}
\email{adurbala@uwsp.edu}

\author[0000-0002-3406-5502]{Adriana Durbala}
\affiliation{Department of Physics and Astronomy, University of Wisconsin – Stevens Point, Stevens Point, WI 54481, USA; adurbala@uwsp.edu}

\author[0000-0001-8518-4862]{Rose A. Finn}
\affiliation{Department of Physics and Astronomy, Siena College, 515 Loudon Rd, Loudonville, NY 12211, USA; rfinn@siena.edu}

\author[0000-0003-0162-1012]{Mary Crone Odekon}
\affiliation{Department of Physics, Skidmore College, Saratoga Springs, NY 12866, USA; mcrone@skidmore.edu}

\author[0000-0001-5334-5166]{Martha P. Haynes}
\affiliation{Cornell Center for Astrophysics and Planetary Science, Cornell University, Ithaca, NY 14853, USA; haynes@astro.cornell.edu}

\author[0000-0002-3144-2501]{Rebecca A. Koopmann}
\affiliation{Department of Physics and Astronomy, Union College, Schenectady, NY 12308, USA; koopmanr@union.edu}

\author[0000-0003-1664-6255]{Aileen A. O'Donoghue}
\affiliation{Department of Physics, St. Lawrence University, 23 Romoda Drive, Canton, NY 13617, USA; aodonoghue@stlawu.edu}

\date{July 2020}

\begin{abstract}
We present an HI-optical catalog of $\sim$ 30,000 galaxies based on the 100\% complete Arecibo Legacy Fast Arecibo L-band Feed Array (ALFALFA) survey combined with data from the Sloan Digital Sky Survey (SDSS). Our goal is to facilitate public use of the completed ALFALFA catalog by providing carefully determined matches to SDSS counterparts, including matches for $\sim$ 12,000 galaxies that do not have SDSS spectra. These identifications can provide a basis for further cross-matching with other surveys using SDSS photometric IDs as a reference point.
We derive absolute magnitudes and stellar masses for each galaxy using optical colors combined with an internal reddening correction designed for small- and intermediate-mass galaxies with active star formation. We also provide measures of stellar masses and star formation rates based on infrared and/or ultraviolet photometry for galaxies that are detected by the Wide-field Infrared Survey Explorer (WISE) and/or the Galaxy Evolution Explorer (GALEX). Finally, we compare the galaxy population in the ALFALFA-SDSS sample with the populations in several other publicly-available galaxy catalogs, and confirm that ALFALFA galaxies typically have lower masses and bluer colors. 
\end{abstract}



\section{Introduction}
\label{sec:intro}
The Arecibo Legacy Fast Arecibo L-band Feed Array (ALFALFA) survey provides HI 21 cm line measurements for $\sim$ 31,500 galaxies over nearly 7000 deg$^2$ on the sky, out to a redshift of about 0.06 \citep{giovanelli05a, haynes18a}. As a ``blind" radio survey, ALFALFA gives an HI-selected view of the low-redshift galaxy population. Primary goals of the survey included determining the HI mass function \citep[e.g.,][]{martin12a, moorman14a, jones18b}, the HI width function \citep{papastergis11a, moorman14a}, and the HI-selected galaxy correlation function \citep{martin12a, papastergis13a}. The HI-selected galaxy population proved to be surprisingly diverse, including many galaxies that still have massive HI disks at low redshift. 

Beyond characterizing the galaxy population in terms of HI alone, ALFALFA is a major resource for comparison with surveys at other wavelengths. Using the 40$\%$ complete ALFALFA survey, ``$\alpha.40$," for example, \citet{huang12a} determined scaling relations between the HI gas fraction and optical and UV colors for a sample of 9417 galaxies with Sloan Digital Sky Survey (SDSS) and Galaxy Evolution Explorer (GALEX) measurements. They found that the specific star formation rate in galaxies with $M_\star < 10^{9.5}$, but not in higher-mass galaxies, is correlated with gas fraction, suggesting that star formation in low-mass galaxies is strongly regulated by HI. Comparing ALFALFA-selected populations with SDSS-selected populations, they found that ALFALFA galaxies have bluer colors, higher star formation rates and specific star formation rates, and lower star formation efficiencies (possibly caused by higher spin parameters).  

The $\alpha.40$ catalog has been the starting point for several follow-up observing programs. For example, the H$\alpha3$ survey \citep{gavazzi13a, fossati13a} is a narrow-band optical imaging follow-up survey of $\sim$ 800 galaxies from ALFALFA in the Local Supercluster and the Coma Supercluster. Among other results, it shows a significant decrease in both HI content and specific star formation rate for galaxies closer to cluster centers, with the outside-in quenching expected from ram-pressure stripping.

Also building from the $\alpha.40$ catalog, the xGASS survey \citep{Catinella18a} combines new and previous HI measurements with GALEX and SDSS measurements to create a gas fraction-limited, stellar mass-selected sample of 1179 galaxies down to a mass of $M_\star = 10^{9} M_\odot $. The related xCOLD GASS survey \citep{Saintonge17} examines molecular gas in 532 galaxies in the same mass range using CO (1-0) observations with the IRAM 30m telescope complemented by CO (2-1) observations with the IRAM 30m and APEX telescopes, HI Arecibo observations, and photometry from SDSS, Wide-field Infrared Survey Explorer (WISE), and GALEX.  They confirmed that the molecular gas fractions depend more strongly on specific star formation rate than on stellar mass.

One particularly promising use of ALFALFA data is in the context of the Baryonic Tully-Fisher Relation (BTFR).  \citet{mcgaugh00a} show that for small galaxies, the combination of gas mass and stellar mass has a surprisingly tight correlation with rotation speed. The BTFR is useful both because it places strong constraints on theories of galaxy formation and evolution and because it enables the measurement of secondary distances and peculiar velocities. \citet{papastergis16a} use $\alpha.40$ to calibrate the BTFR for gas-dominated galaxies. 

A primary purpose of the current paper is to facilitate public use of the recently completed 100\% ALFALFA survey in conjunction with observations at other wavelengths. First, we provide SDSS identifications for nearly all (29,418 out of 31,501) ALFALFA galaxies, including $\sim$ 12,000 that are relatively difficult to identify because they do not have SDSS spectroscopy. The SDSS identifications can provide a basis for further cross-matching with other surveys using SDSS photometric IDs as a reference point. We derive absolute magnitudes and stellar masses for each galaxy using optical colors combined with an internal reddening correction designed for small- and intermediate-mass galaxies with active star formation. We also provide measures of stellar masses and star formation rates based on infrared and/or ultraviolet photometry for galaxies that are detected by WISE and/or GALEX. In this way, the final catalog puts the HI information from ALFALFA in the context of each galaxy's stellar content.

This paper is organized as follows. Section 2 describes the identification of HI sources, the assignment of SDSS optical counterparts, optical extinction corrections, and stellar masses calculations based on optical colors. Section 3 describes stellar masses and star formation rate calculations using infrared and ultraviolet photometry. Section 4 presents the ALFALFA-SDSS catalog.
Section 5 places the ALFALFA-SDSS catalog in the context of three other galaxy catalogs that include stellar mass and/or star formation rate, showing the overall differences between the ALFALFA-SDSS galaxy population and populations with other selection criteria. Section 6 summarizes the content and scope of the catalog.

Throughout this paper we adopt the cosmological parameters $\Omega_{m}$ = 0.3, $\Omega_{\Lambda}$ = 0.7, and H$_0$ = 70 km s$^{-1}$ Mpc$^{-1}$.

\section{SDSS Counterparts and Optical Properties}\label{sec:sdsscounterparts}

Exploiting the large collecting area of the Arecibo 305~m antenna and the 7-beam ALFA radio camera, ALFALFA mapped $\sim$6,600 deg$^2$ of high galactic latitude sky in spectral line mode, covering a 100 MHz bandwidth corresponding to 2000 $<$ c$z <$ 18000 km s$^{-1}$ sampled as 4096 spectral channels, yielding a resolution of 5.5 \kms at $z \sim$ = 0 before smoothing \citep{giovanelli05a,haynes11a}. As described in Appendix A of \citet{haynes11a}, multiple fixed-position drift scan crossings were combined to construct three-dimensional (position-position-velocity) grids. Source identification was performed by applying a Fourier domain-based matched filter algorithm to the processed spectral grids \citep{saintonge07a}. Source extraction and parameter measurement was performed by interactive analysis of each candidate detection, allowing for localized baseline fitting and parameter extraction.

The size of a single ALFALFA beam is 3$^\prime$.8~$\times$~3\arcmin.3. The grids are constructed on a spatial grid sampled at 1\arcmin~$\times$~1\arcmin~ to which a Gaussian weight function of 2\arcmin~ is applied. This reduces the spatial resolution of the grid to $\sim$ 4\arcmin.3$\times$~3\arcmin.8. As discussed in Section 5 of \citet{giovanelli07a}, the pointing accuracy of the extracted HI sources is limited by the resolution of each ALFA beam, the signal to noise ratio (SNR) of the HI emission, and the quality of the telescope's pointing. The latter has been accounted for by fitting offsets between peaks in the continuum sources associated with each spectral grid and the positions of radio sources catalogued by the Northern VLA Sky Survey \citep[NVSS;][]{condon98}. Examination of both the scatter in the continuum offsets and the difference in positions between the HI sources and their likely optical counterparts (see next section) show that the positional accuracy of high SNR ($>$10) sources is 15-18\arcsec, but can be much larger, even exceeding 1\arcmin~ at lower SNR \citep{giovanelli07a, kent08a}. 

\subsection{Identification of Optical Counterparts} 

In addition to the parameters of the HI emission, the ALFALFA catalog includes the position of the ``most probable'' optical counterpart. The process by which optical counterparts are assigned to the HI sources is discussed in Appendix A and especially Figures 8 and 9 of \citet{haynes11a}. Essentially, the individual interactively performing the source parameter extraction examined, at the same time, several digital extragalactic source catalogs and imaging databases to search for likely stellar counterparts to the HI emission. The search box for coincidence was adjusted to take into account the SNR as discussed above, with larger regions searched for sources of lower SNR. Within the positional error box of the vast majority of HI sources lies a star-forming galaxy which was deemed the most probable optical counterpart. Where the optical identification was less obvious, a best candidate was identified based on proximity to the HI centroid, morphological appearance as a star-forming disk galaxy, and where available, known and coincident radial velocity. The latter information in particular was used as an additional criterion for the identification of a separate category of HI sources (HI code = 2, the ``priors'') with SNR below the threshold of 6.5 applied to the high quality detections (HI code = 1). 

The ALFALFA source catalog presented in \citet{haynes18a} contains 31501 sources, of which 25433 (80\%) are high quality and 6068 are lower SNR ``priors''.  As discussed by \citet{haynes18a} and \citet{leisman17a}, nearly all of the high quality ALFALFA sources can be identified with a likely optical counterpart. In fact, of the high SNR sources, only 344 (1.4\%) are not assigned a probable optical counterpart and most of those are probably tidal debris. For the 25089 high quality HI sources with identified optical counterparts, the median separation of the position of the identified optical counterpart from the HI centroid is 19\arcsec.9; the mean is 23\arcsec.6 with a standard deviation of 16\arcsec.8. Some of this separation value is due to uncertainties in the optical positions recorded in the ALFALFA database, generally estimated to be $<$3\arcsec.

In Table 3 of \citet{haynes11a}, we presented  galaxies in the 40\% ALFALFA HI catalog, along with a cross-match to SDSS DR7. The cross identification was performed as part of the data reduction process, whereby the user marked the SDSS counterpart interactively by visual inspection of the SDSS image served by the NAVIGATOR tool, as part of the standard data reduction process. Later SDSS data releases presented new cross identifications with new photometric solutions, so the process of cross-identification was reinitiated for the full ALFALFA catalog in 2017. In the latter case, a more automated approach was adopted, but it made use of the positional matches made during the earlier visual inspection.

For the final ALFALFA-SDSS crossmatch reported here (ALFALFA-SDSS), a more automated approach was performed using the SDSS Cross ID tool\footnote{http://skyserver.sdss3.org/public/en/tools/crossid/crossid.aspx.}. 
 Matches were sought within a search radius of 0\arcmin.1 around the center of the optical counterpart identified in the ALFALFA database. Galaxies with anomalous magnitudes were inspected individually. Many of these were galaxies with dust lanes, bright HII regions, or superposed stars, where there were multiple photometric sources. In some cases, it was possible to reassign an appropriate SDSS source for the galaxy as a whole.  In other cases this was not possible.
 
Most galaxies with a clear SDSS counterpart are assigned an optical photometry flag of ``1'' (28057 objects). Galaxies with a clear SDSS counterpart but with large photometric uncertainties (as described in the section on optical photometry below) are assigned the flag ``2'' (1361 objects). Galaxies with no clear SDSS counterparts are assigned the flag ``0'' if they were outside the SDSS footprint (1296 objects) and ``3'' otherwise (787 objects). 

Distances to each galaxy are estimated using the process described in
\citet{haynes18a}. 

\subsection{Optical Photometry and Extinction}\label{sec:optphot}

Following the recommendations on the SDSS website, we use SDSS {\it cmodel mags} to calculate galaxy absolute magnitudes and {\it model mags} for colors. Galaxies with g- or i-band errors greater than 0.05 were assigned a photometry flag of ``2" and excluded from color-magnitude diagrams and stellar mass calculations based on optical magnitudes.  Excluding these galaxies cuts out the majority of objects with anomalous colors that suggest magnitude uncertainties beyond the formal errors calculated by the SDSS pipeline. Inspection of individual galaxies indicates that several different factors contribute to the anomalous magnitudes, including contamination by nearby stars and shredding of large galaxies. We also note that the SDSS provides a ``clean'' parameter, but using this as a flag preferentially excludes the small, blue galaxies that dominate our sample; a full 43\% of our galaxies would be left out, most of which look normal otherwise. Our flag based on magnitude uncertainties leaves out only 4\% of the sample. 

We correct optical photometry for foreground extinction by the Milky Way using the $E(B-V)$ map of \citet{schlegel1998a} with the $R_V=3.1$ reddening curve of \citet{schlafly11a}. We do not apply the additional  14\% recalibration to lower values for extinction suggested by \citet{schlafly11a}, because not all authors agree the calibration should be lower. For example, the \citet{planck14a} find the values in \citet{schlegel1998a} to be too low (by about 8\%) rather than too high. We note that for the typical values of $E(B-V)$ in our sample, around 0.06, a shift of about 10\% in $E(B-V)$ corresponds to a change in magnitude of only 0.02 in the g band and 0.01 in the i band. Indeed, the scatter in the results from different methods is larger than the systematic shift. For the purposes of error propagation, we adopt an uncertainty of 20\% in the values of g- and i-band galactic extinction corrections (in magnitudes), and an uncertainty of 0.02 for g-i color \citep[see][]{green15a, green19a}.

Galactic extinction values provided by the SDSS for Data Release 15 are the same as our chosen values; they also use the conversions from E(B-V), but not the 14\% recalibration, of \citet{schlafly11a}. (The SDSS pipeline does not apply galactic extinction corrections automatically; this is left to the user.)

\begin{figure*}[htp]
    \centering
    \includegraphics[width=.9\textwidth]{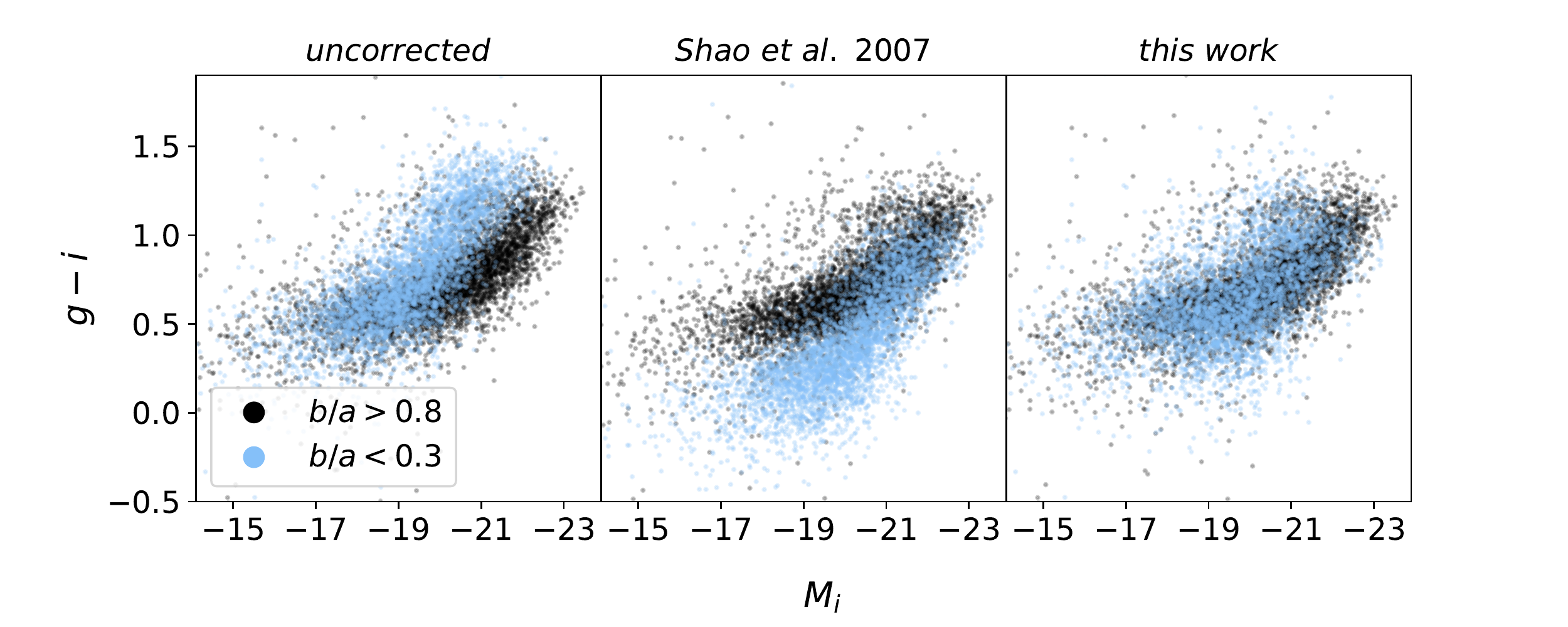}
    \caption{The effect of the internal extinction correction on the color-magnitude diagram for the cases of (left) no correction, (center) the correction from \citet{shao07}, and (right) the correction adopted in this work (see text for details).
      Each panel compares galaxies that have highly inclined disks ($b/a<0.3$, light blue) with galaxies that appear nearly face-on ($b/a>0.8$, black).  We expect these two groups to have similar colors and magnitudes with an appropriate extinction correction.}
    \label{fig:extinction}
\end{figure*}

We also correct optical photometry for extinction internal to each galaxy. A simple and standard way of doing this is to estimate the extinction in magnitudes using $A_\lambda = \gamma_\lambda log10(a/b)$, where $a/b$ is the axial ratio obtained from $expAB\_r$ in SDSS, and $\gamma_\lambda$ is a constant for each filter \citep[see][]{giovanelli94a, shao07}. However, if we use the same $\gamma$ for all the galaxies in our sample, this method clearly over-corrects the photometry for fainter galaxies.  As shown in Figure \ref{fig:extinction}, the ``corrected" $g-i$ colors for edge-on faint galaxies, for example, are bluer than those for face-on galaxies by about half a magnitude. Indeed, there has long been evidence that there is less internal extinction in less luminous galaxies \citep[e.g.][]{tully98a}. There is also increasing evidence for still more complicated extinction effects, including a nonlinear dependence of $A_\lambda$ on $log10(a/b)$, and differences in extinction based on additional parameters, including bulge-to-disk ratios, colors, gas content, and surface brightness \citep{masters10a, devour16a, kourkchi19a}.  

Our approach here is to provide a simple correction that captures the observed overall dependence of extinction on absolute magnitude for galaxies in the blue cloud.  We use a value of $\gamma$ that is zero for absolute magnitudes fainter than $-17$, and that changes linearly for brighter magnitudes according to
\begin{equation}
\begin{aligned}
  \gamma_g & = -0.35M_g - 5.95, M_g < -17 \\
  \gamma_i & = -0.15M_i - 2.55, M_i < -17
\end{aligned}
\end{equation}
with an uncertainty of $\pm 0.3$  in $\gamma$. 
This correction is designed to be consistent with the results of \citet{masters10a} and \citet{devour16a} for the case of star-forming galaxies with small and intermediate magnitudes. 
We do not recommend using it for passive (red sequence) galaxies or for very massive galaxies, both of which show evidence for less extinction than that given by these equations.    
Figure \ref{fig:extinction} illustrates the effect of the adopted internal extinction correction on the galaxy color-magnitude diagram, compared with using either no correction at all, or the correction from \citet{shao07}.  Each panel compares galaxies that have highly inclined disks ($b/a<0.3$) with galaxies that appear nearly face-on ($b/a>0.8$).  We expect these two groups to have similar colors and magnitudes with an appropriate extinction correction. 

\subsection{Derivation of Stellar Masses using SDSS Photometry}\label{sec:stellmass}

The stellar mass of a galaxy is a fundamental property that, when combined with information about whether the galaxy is star-forming or quenched, can closely predict many other galaxy properties, including  color, shape, and metallicity \citep[e.g.,][]{balogh04a, 
brinchmann04a, blanton05a, mouhcine07a, brough13a}. 
Characterizing galaxies in terms of their stellar mass is powerful and intuitive, and provides a way to compare results across different types of galaxy surveys. 
Stellar mass can also be used in conjunction with gas mass to determine distances according to the Tully-Fisher relation \citep{tully77a}.  In the context of the ALFALFA-SDSS catalog, we are particularly interested in the {\it baryonic} Tully-Fisher relation \citep{mcgaugh00a}, which improves the simple Tully-Fisher relation for small galaxies by including gas mass as well as stellar mass.  This gives us an additional motivation for characterizing the ALFALFA galaxies in terms of their stellar mass.

Stellar masses can be estimated by comparing spectra and/or or broadband photometry to stellar population synthesis models \citep{kauffmann03, chen12a, maraston13a}.  The GALEX-SDSS-WISE Legacy Catalog 2 \citep[GSWLC-2;][]{salim16a, salim18a}, for example, uses UV/optical/IR SED fitting to calculate stellar masses and star formation rates. 
 
For the purposes of providing stellar masses for the ALFALFA galaxy population as a whole, however, we cannot take advantage of the values provided by these catalogs 
because of the lack of sufficient overlap between them and our survey volume. This is especially true for the part of ALFALFA in the ``fall sky" (galactic southern hemisphere) not covered by the SDSS spectroscopic surveys.  Furthermore, many existing catalogs that provide stellar masses for large samples of galaxies \citep[e.g.][]{blanton11a,salim16a} are incomplete at the low stellar masses typical of ALFALFA galaxies.

\begin{figure*}[htp]
    \centering
    \includegraphics[width=0.48\textwidth]{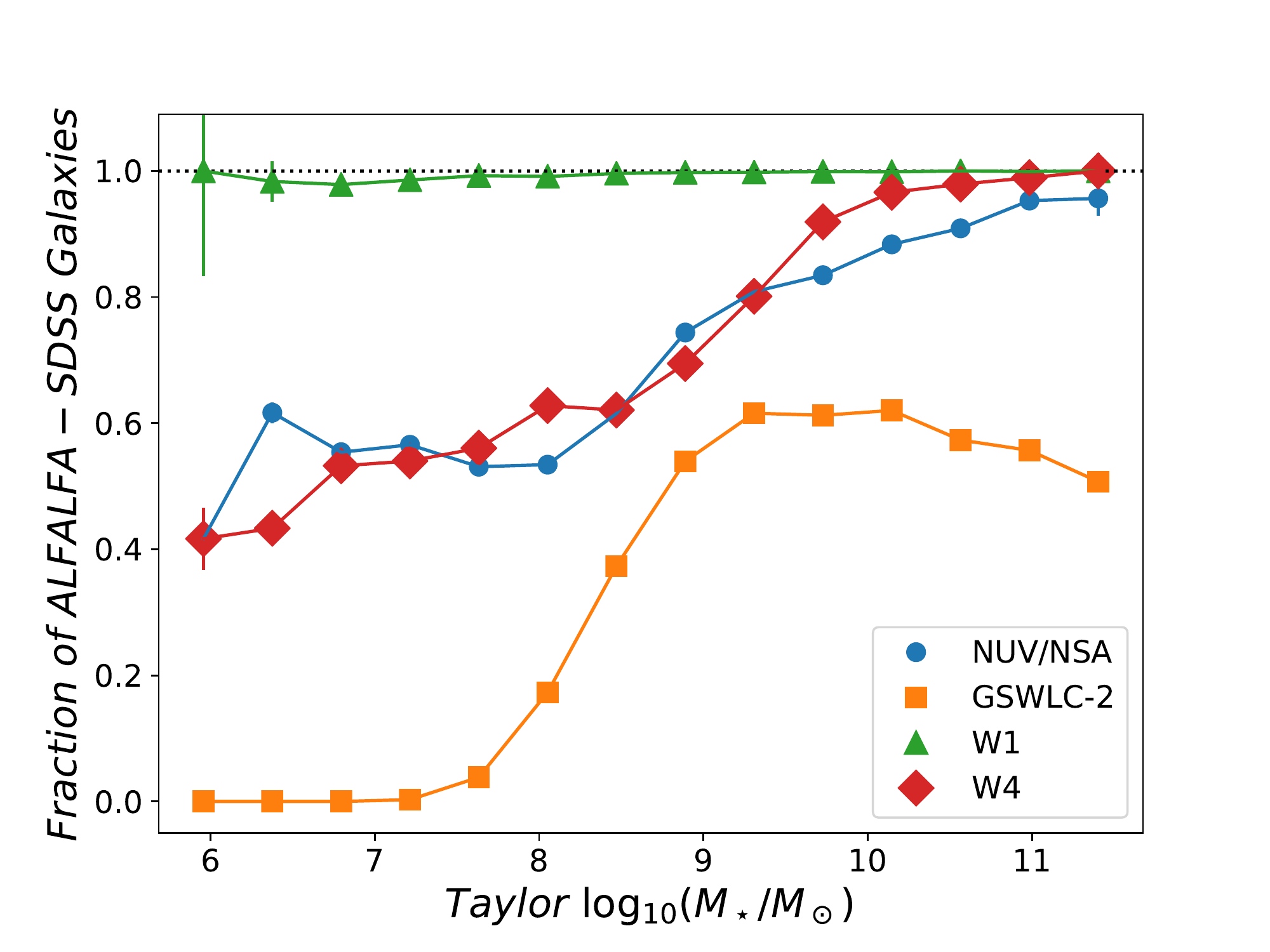}
    \includegraphics[width=0.48\textwidth]{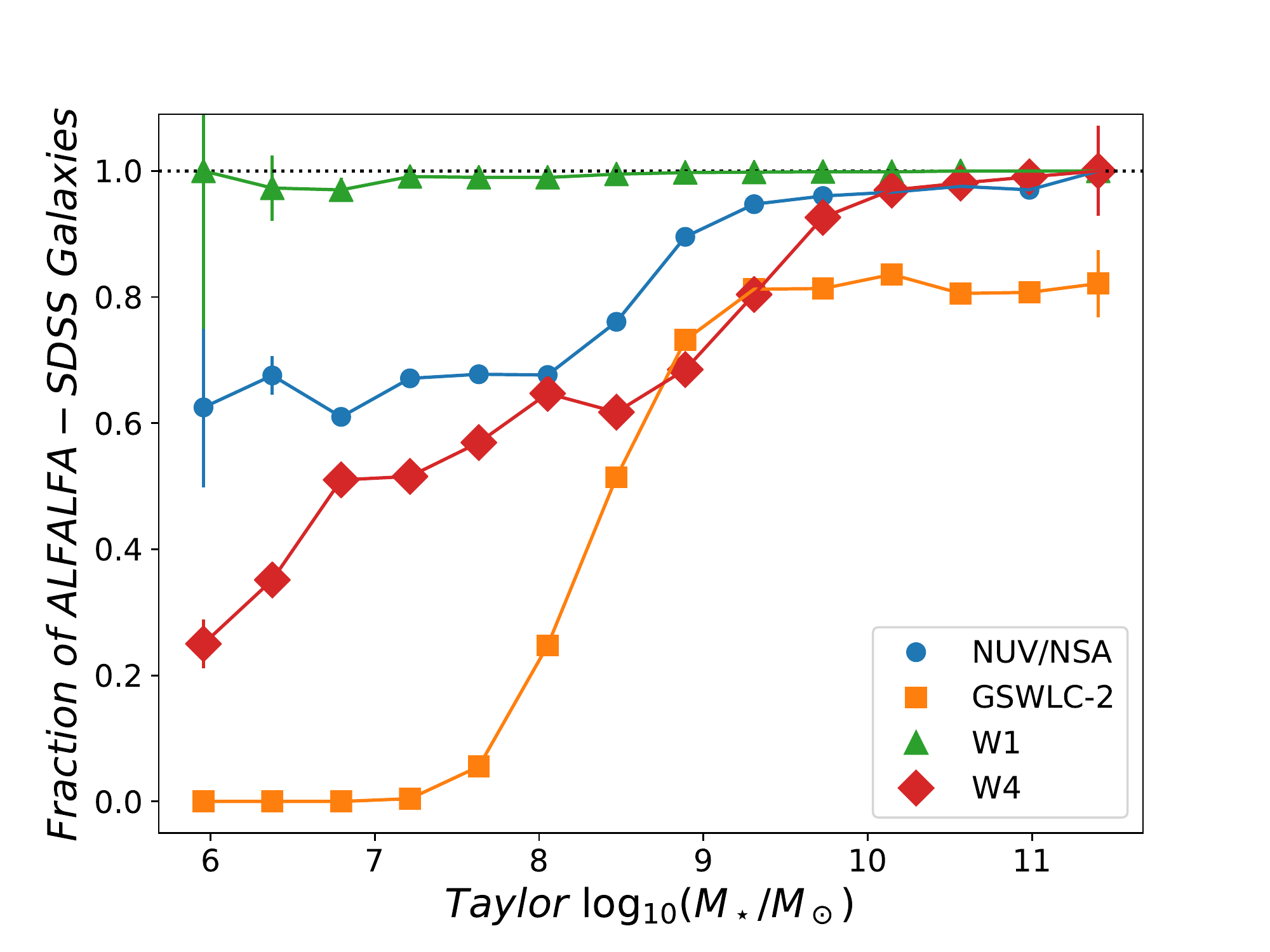}
    \caption{(Left) Fraction of ALFALFA-SDSS galaxies included in NSA (blue circles), GSWLC-2 (orange squares), and unWISE W1 (green triangles) and W4 (red diamonds) surveys as a function of stellar mass.  We use the full survey areas for this comparison. (Right) Same as left panel, but for the survey overlap regions only.}
    \label{fig:detection-fraction}
\end{figure*}

To illustrate these coverage and completeness issues, we show the fraction of ALFALFA-SDSS galaxies detected by several surveys in 
Figure \ref{fig:detection-fraction}.  In the left panel, we show the fraction of \aone \ galaxies included in the GSWLC-2 (orange squares) when we compare the full regions covered by each survey.  To estimate completeness, we limit the comparison to a region where the surveys overlap: $140\degrees <$  R.A. $< 230\degrees$, $0\degrees <$ Dec. $< 35\degrees$, and $z<0.05$.  In the right panel of Figure \ref{fig:detection-fraction} we see that even in the overlap region, the GSWLC-2 misses a large fraction of \aone \ galaxies with $\log_{10}(M_\star/M_\odot) < 9$.
With an eye toward calculating star-formation rates as well as stellar masses (Sections \ref{sec:IRsfr}, \ref{sec:IRstellmass}), we also show the fraction of \aone \ galaxies that have GALEX NUV fluxes (blue circles) reported in the NASA-Sloan Atlas  \citep[NSA;][]{blanton11a}, and the fraction detected at W4 (22\micron) in the unWISE catalog \citep[red diamonds; ][]{lang14,lang16}.  While these surveys include a higher fraction of \aone \ galaxies than the GSWLC-2, they still miss the majority of low-mass galaxies.
Almost all ALFALFA-SDSS galaxies seem to be detected in unWISE W1 (3.4\micron) (green triangles), and we will utilize this to calculate an IR-based stellar mass in Section \ref{sec:IRstellmass}.  

The incomplete sampling of ALFALFA galaxies in existing catalogs drives us to calculate independent stellar mass estimates from SDSS photometry.  
As an optically-based measure of stellar mass, we adopt the method of \citet{taylor11a}, which can be used for all the galaxies in the matched ALFALFA-SDSS catalog, including those without optical spectra. This simple prescription is based on optical color and magnitude according to: 
\begin{equation}
\log M_{\star}/L_{i}=-0.68+0.70(g-i)
\end{equation}
\citet{taylor11a} show that their method agrees with more complicated ones. 
We denote stellar mass determined using the Taylor method as M$_{\star, Taylor}$.

About 40\% of the galaxies in the ALFALFA-SDSS catalog have stellar masses from the GSWLC-2, and we compare our stellar mass estimates with theirs in the top left panel of Figure \ref{fig:mstar-gswlc}.  The two mass estimates are tightly correlated, but the M$_{\star, Taylor}$ estimates tend to be slightly below the GSWLC-2 masses, and the offset increases with stellar mass. We provide a translation between the two mass estimates by fitting a linear relationship: 
\begin{equation}
\begin{split}
\scriptsize
\rm
    \log_{10}(M_\star/M_\odot)_{GSWLC-2} = 1.05 \log_{10}(M_\star/M_\odot)_{Taylor}\\
     -0.37
\end{split}
\end{equation}
When we use this fit to correct the M$_{\star, Taylor}$ estimates, we are able to remove the dominant systematic offsets. This is borne out in the bottom left figure, which shows the residuals between the GSWLC-2 and corrected Taylor stellar masses. 
The dispersion of the residuals is only 0.11~dex.

In the ALFALFA-SDSS catalog (Section ~\ref{sec:catalog}), we include the (uncorrected) values for M$_{\star, Taylor}$ as well as the GSWLC-2 values for the $\sim 40\%$ of galaxies for which they are available.

\section{Infrared and Ultraviolet Properties}\label{sec:wisecounterparts}

\subsection{Derivation of Stellar Masses using unWISE Photometry}\label{sec:IRstellmass}

Infrared photometry provides an additional method for estimating stellar masses, and is available for a large fraction of the galaxies in the ALFALFA-SDSS catalog thanks to the all-sky coverage of NASA's Wide-field Infrared Survey Explorer \citep[WISE\footnote{The Wide-field Infrared Survey Explorer (WISE)
All-Sky Data Release is available at http://wise2.ipac.caltech.edu/docs/release/allsky/ }; ][]{wright10}.  WISE mapped the sky in four infrared bands: $3.4 \mu m$ (W1), $4.6 \mu m$ (W2), $12 \mu m$ (W3), and $22 \mu m$ (W4) with an angular resolution of 6.1", 6.4", 6.5", and 12.0". 
The unWISE catalog is derived from a reprocessing of WISE imaging \citep{lang14} and has two big advantages over the existing WISE  catalog (AllWISE): deeper imaging and improved modeling of crowded regions. It detects sources at a 5$\sigma$ level to about $\sim$ 0.7 magnitudes fainter than the AllWISE catalog, doubling the number of galaxies detected between redshifts  $0 < z < 1$. In the end, unWISE offers more accurate photometry for extended sources than the AllWISE catalog, and we therefore choose to use it in our study for identification of infrared counterparts. 

There are are two databases associated with unWISE.  One is the unWISE Catalog, a point-source catalog with sources identified by a unique unWISE ID \citep{schlafly19}.  The other is the unWISE/SDSS Forced Photometry catalog, which takes SDSS sources and shapes, and fits for unWISE fluxes that best match the unWISE images \citep{lang16}.  ALFALFA sources were matched to unWISE sources by SDSS objID number in this catalog (D. Lang, private communication). This yields 29,088 ALFALFA sources with an unWISE match. 



\begin{figure*}[htp]
    \centering
    \includegraphics[width=.9\textwidth]{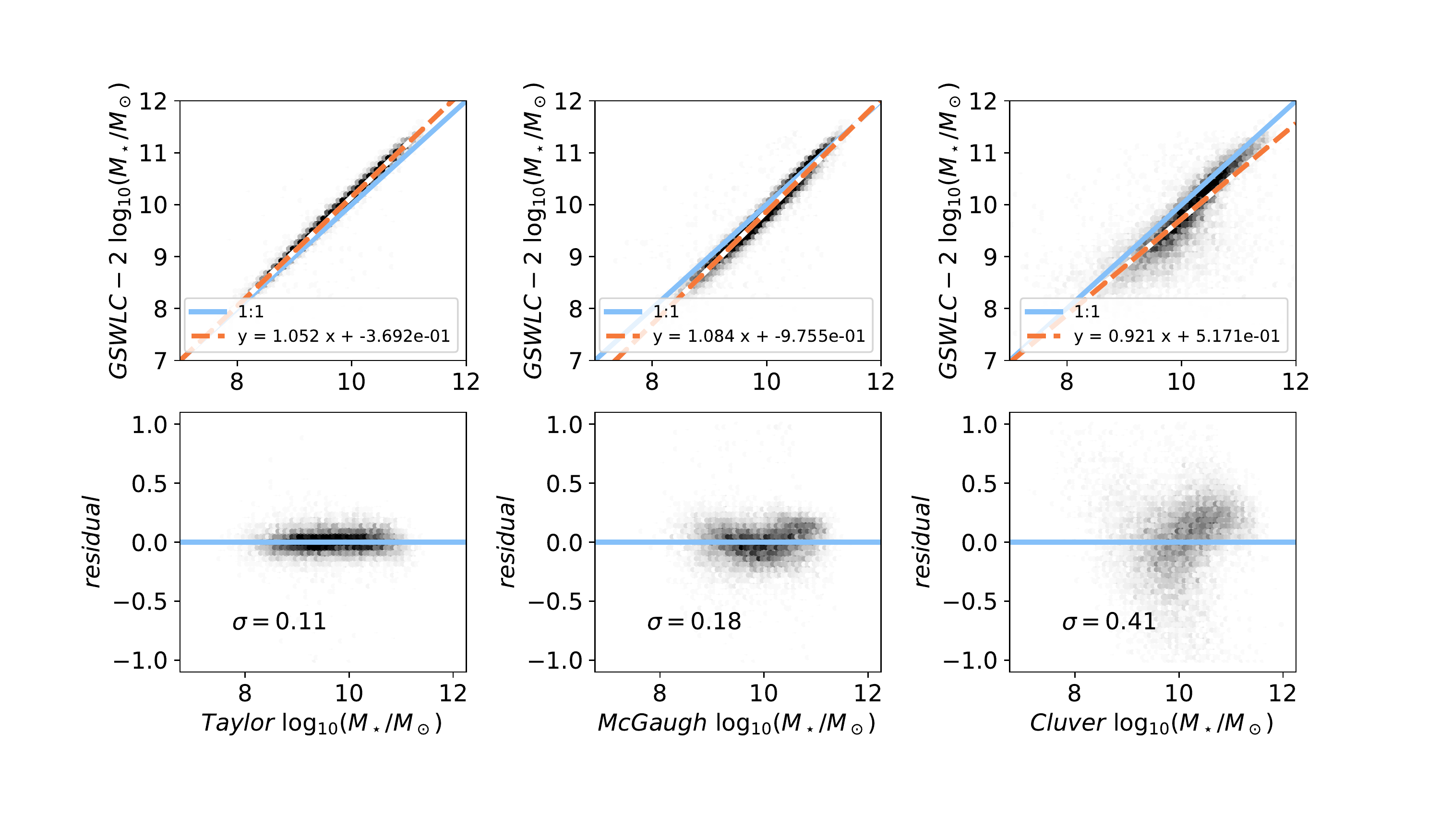}
    \caption{Top Panels: GSWLC-2 stellar masses versus (left) stellar mass estimates from SDSS $g-i$ color \citep{taylor11a}, (center) unWISE W1 \citep{mcgaugh15a}, and unWISE $W1-W2$ color \citep{cluver14}. The light blue line shows a 1-to-1 relation, and the orange dashed line shows a linear fit.  Bottom Panels: Residuals from the fitted relation for each mass estimate.  For both the \citet{taylor11a} and \citet{mcgaugh15a} mass estimates, we are able to remove systematic offsets with respect to stellar mass.}
    \label{fig:mstar-gswlc}
\end{figure*}

We consider two methods of estimating stellar masses using infrared photometry from unWISE. \citet{mcgaugh15a} calculate stellar mass using the WISE W1 band, which is dominated by light from old stars \citep{jarrett13}: 
\begin{equation}
M_\star/L_{W1} = 0.45 M_{\sun}/L_{\sun}
\end{equation}
We denote stellar masses determined using this method as M$_{\star, McGaugh}$.
The center column of Figure \ref{fig:mstar-gswlc} compares the GSWLC-2 masses with the M$_{\star, McGaugh}$ estimates. We are able to remove the dominant systematic offsets between the two mass estimates using a linear fit.  However, the scatter between the McGaugh and GSWLC-2 masses is systematically larger than for the Taylor-GSWLC-2 masses. As discussed in \citet{salim16a}, these differences may be related to uncertainties in the contribution from post-main sequence stars to W1 flux. The best fit conversion between the \citet{mcgaugh15a} and GSWLC-2 masses is:
\begin{equation}
\label{eqn:mcgaugh_mstar}
\begin{split}
\rm
\scriptsize
\log_{10}(M_\star/M_\odot)_{GSWLC-2} = 1.084
    \log_{10}(M_\star/M_\odot)_{McGaugh} \\ -0.9755 \\
\end{split}
\end{equation}

for $$8 <  \log_{10}(M_\star/M_\odot)_{McGaugh} < 11$$ 

Another measure of stellar mass, derived by \citet{cluver14}, combines the \textit{WISE} W1 luminosity and the \textit{WISE} W1-W2 color:
\begin{equation}
\log_{10} (M_{\star}/L_{W1}) = -2.54 (W_{3.4\mu m}-W_{4.6\mu m}) -0.17.
\end{equation}
Here, $W_{3.4\mu m}-W_{4.6 \mu m}$ is the rest-frame color of the source and $L_{W1} (L_{\sun})=10^{-0.4(M-M_{\Sun})}$, where $M_{\Sun}=3.24$ and M is the absolute magnitude of the source in W1 band. 
We denote stellar masses determined using this alternative method as M$_{\star, Cluver}$.
We compare this stellar mass estimate with the GSWLC-2 mass in the right column of Figure \ref{fig:mstar-gswlc}. The \citet{cluver14} stellar masses are reasonably consistent for $\log_{10}(M_\star) > 10$, but the scatter and offset are large for lower mass galaxies. Therefore, we do not report a best-fit relation, and we do not include the \citet{cluver14} stellar masses in the catalog (Section~\ref{sec:catalog}).

\subsection{Derivation of Star-Formation Rates using IR and UV photometry}\label{sec:IRsfr}

Several well-characterized indicators are used to trace recent star formation, including the direct detection of UV radiation from massive stars, and the infrared emission from dust grains that absorb some the UV light and re-radiate it in the infrared.  When both UV and infrared fluxes are available, the combination provides a SFR tracer that is robust against extinction.  The combination of WISE and GALEX makes these measurements possible for large samples of nearby galaxies \citep[e.g.,][]{salim18a, leroy19a}.  


We provide multiple measures of total star-formation rate for galaxies that are detected by WISE and/or GALEX.
First, we calculate star-formation rates using the 22\micron \ flux and the conversions from \citet {kennicutt12a} based on calibrations from \citet{rieke09a}. A total of 23,895 galaxies have a detection in W4 (we require W4$>$0). 
We convert the unWISE W4 magnitudes from Vega to AB by adding 6.620 \citep{jarrett11}, and then to Janskys using a flux zeropoint of 3631~Jy.  We compute $\nu L_\nu$ by multiplying the flux in Jy by the frequency at 22\micron \ and by $4 \pi D^2$, where $D$ is the flow-corrected distance from \citet{haynes18a} (see Table \ref{tab:catalog1}).  According to \citet{kennicutt12a}, the SFR is then 
\begin{equation}
    \log_{10}(SFR_{22}) = \log_{10}(\nu L_{\nu_{22}}) - 42.69.
\end{equation}
We denote star formation rates determined this way as $\log_{10}(SFR_{22})$.

For galaxies that have GALEX near-ultraviolet fluxes in the NASA-Sloan Atlas \citep{blanton11a}, we calculate the NUV star-formation rates \citep{kennicutt12a,hao11a, murphy11a} and corrected NUV star-formation rates \citep{kennicutt12a,hao11a} respectively as
\begin{equation}
\log_{10} (SFR_{NUV}) = \log_{10} (\nu L_\nu) - 43.17
\end{equation}
and
\begin{equation}
\log_{10} (SFR_{NUV_{corr}}) = \log_{10} (\nu L_{\nu_{corrected}}) - 43.17.
\end{equation}
where the corrected NUV spectral energy density $\nu L_{\nu_{corrected}}$ is the sum $\nu L_\nu(NUV) + 2.26 \nu L_\nu(22\micron$).
A total of 22848 ALFALFA sources have NUV detections. 

\begin{figure*}[htp]
    \centering
    \includegraphics[width=0.9\textwidth]{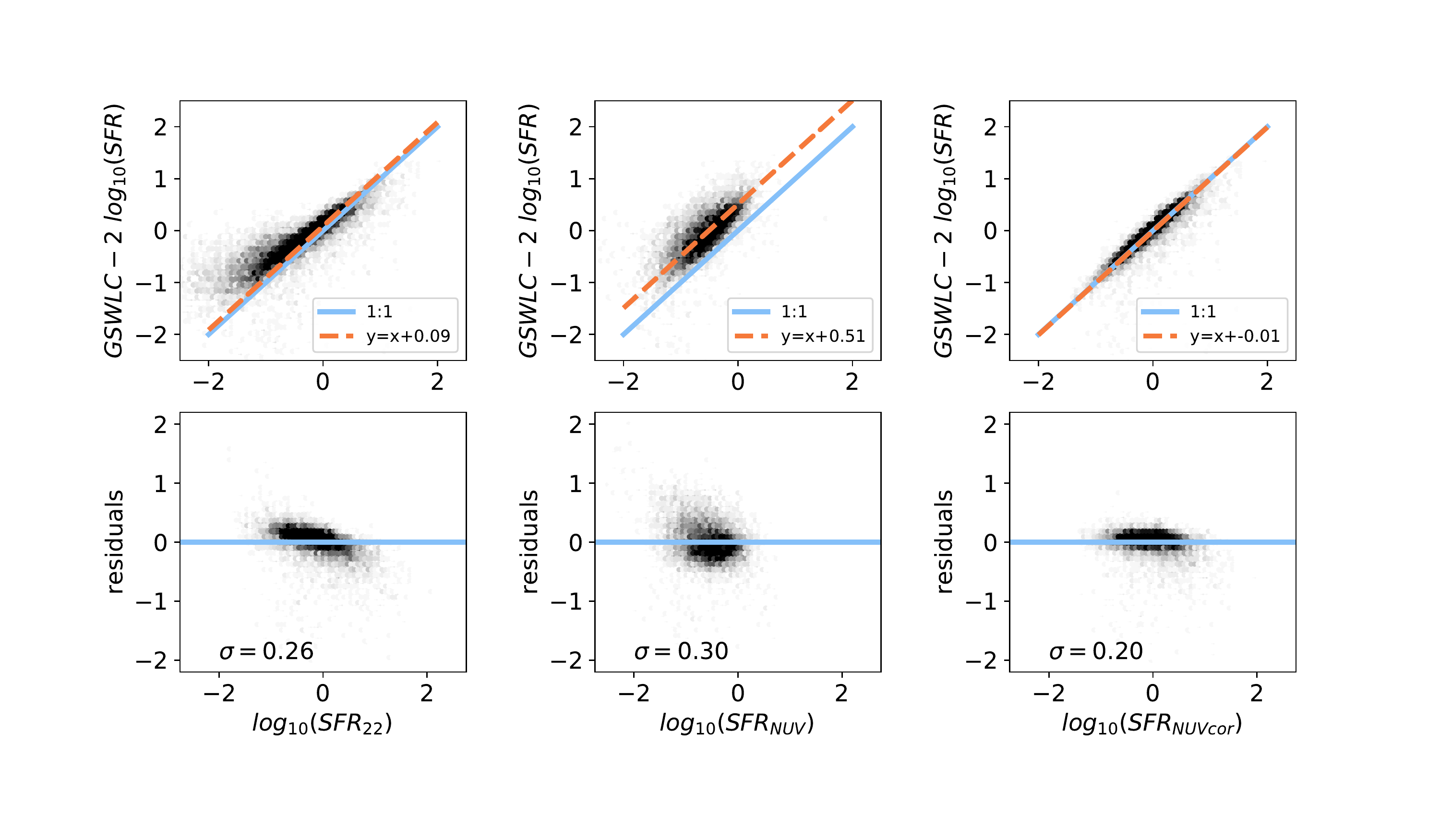}
    \caption{Top Panels: GSWLC-2 SFR versus estimates from (left) $WISE$ W4, (center) GALEX near-UV, and (right) corrected near-UV.  The light blue line shows the one-to-one relation, and the orange dashed line has a slope of one but an intercept that is the best-fit zeropoint offset between the two indicators.  Bottom panels: Residuals between the two SFR indicators after correcting for the zeropoint offset.  The corrected NUV SFR provides the most reliable estimate of SFR and should be used when available.}
    \label{fig:fit_sfr}
\end{figure*}

In Figure \ref{fig:fit_sfr}, we compare the three measures of SFR with the values from the GSWLC-2 \citep{salim16a,salim18a}.  The GSWLC-2 SFRs are based on SED fitting from  UV/optical photometry jointly with the mid-IR flux from 22\micron, or 12\micron\ if 22\micron\ is not detected \citep{salim18a}.  In the left column, we show the GSWLC-2 SFR versus the $WISE$ 22\micron \ SFR.  The two measures of SFR are in good agreement for $log_{10}(SFR) > 0$ but the 22\micron \ SFRs fall below the GSWLC-2 values for lower SFRs and lower mass galaxies.  This is expected due to the lower metallicity and dust in lower-mass galaxies, which results in less extinction.
We find an average offset of 0.09, but this is due mostly to the offsets observed in lower SFR galaxies.  We don't fit a linear relationship because this would make the $SFR_{22} > 0$ values inaccurate.  Instead, we prefer to use the 22\micron \ SFR with the caveat that they will underestimate the true SFR for low-mass, low SFR galaxies.  The NUV SFR (middle column) underestimates the total SFR by an average of 0.51 due to dust absorption, which again is a strong function of stellar mass.  When the NUV flux is corrected for emission that is absorbed and re-radiated in the IR, the inferred SFR is much closer to GSWLC-2 values (right column). Therefore, when both NUV and IR are available, the corrected NUV SFR should be used.  Otherwise, the 22\micron \ SFR is the next best option. NUV alone is the least reliable, and we do not include it in our catalog.

\section{The  ALFALFA-SDSS Catalog}\label{sec:catalog}

 Table 1 presents the ALFALFA 100\% catalog including basic SDSS properties of cross-matched galaxies.
The table is organized as follows:

\begin{enumerate}
\item Column 1---AGC number (entry number in the AGC catalog)
\item Column 2---photometry flag. 0: outside the SDSS footprint, 1: SDSS photometry with uncertainties less than 0.05 in g and i (\textit{good photometry}), 2: SDSS photometry with uncertainties greater than 0.05 in g and/or i (\textit{bad photometry}), 3: no SDSS counterpart identified, despite being within the SDSS footprint
\item Column 3---SDSS DR15 Object ID of the optical counterpart
\item Column 4---right ascension (J2000) of the optical counterpart or HI centroid, if no optical counterpart has been identified
\item Column 5---declination (J2000) of the optical counterpart or HI centroid, if no optical counterpart has been identified
\item Column 6---heliocentric velocity (cz) of the HI profile midpoint  in \kms 
\item Column 7---distance in Mpc estimated as described in
\citet{haynes18a}
\item Column 8--- uncertainty in distance from \citet{haynes18a}
\item Column 9---g-band Galactic extinction in mag, as described in section 2.2
\item Column 10---i-band Galactic extinction in mag, as described in section 2.2
\item Column 11---$expAB\_r$ axial ratio b/a in r band from SDSS
\item Column 12---uncertainty in $expAB\_r$ axial ratio b/a in r band from SDSS
\item Column 13---SDSS i-band cmodel magnitude 
\item Column 15---uncertainty in i-band cmodel magnitude from SDSS

\end{enumerate}
We include additional information for each galaxy, including all the  photometric measurements from SDSS needed to calculate the derived optical properties in a table at http://egg.astro.cornell.edu/alfalfa/data/index.php. 

\begin{table*}
\begin{center}
\scriptsize
\setlength\tabcolsep{3.0pt} 
\caption{Basic Optical Properties of Cross-listed objects in the ALFALFA-SDSS Catalog.\label{tab:catalog1}  This table is available in its entirety online.} 
\begin{tabular}{|c|c|c|c|c|c|c|c|c|c|c|c|c|c|}
\hline 
\toprule 
AGC &	Flag &	SDSS objID  & RA &	DEC &	V$_{helio}$ &	D &	$\sigma_D$  &	Ext$_g$	& Ext$_i$	& expAB$_r$  & $\sigma_{expAB_r}$ &	cmodel$_i$ & $\sigma_{cmodel_i}$ \\ 
& & & J2000 & J2000 & $km~s^{-1}$ & Mpc & Mpc & mag & mag & & & mag & mag \\ 
(1) & (2) & (3) & (4) & (5) & (6) & (7) & (8) & (9) & (10) & (11) & (12) & (13) & (14)  \\ 
\midrule 
\hline 
1 & 1 & 1237679455462228052 &  0.656670& 16.65222 & 5839 & 82.8 & 2.2 &  0.11 & 0.06& 0.77& 0.01& 15.07 &0.01\\ 
3 & 1 & 1237679502171701451 &  0.692920& 18.88583 & 7883 & 107.6 & 2.3 &  0.11 & 0.06& 0.47& 0.01& 13.08 &0.01\\ 
4 & 1 & 1237678660887445652 &  0.737080&  4.20889 & 8621 & 118.0 & 2.3 &  0.09 & 0.05& 0.64& 0.01& 13.88 &0.01\\ 
6 & 1 & 1237679476933722181 &  0.790420& 21.95972 & 6561 & 88.8 & 2.2 &  0.15 & 0.08& 0.73& 0.01& 13.52 &0.01\\ 
7 & 1 & 1237656496724639763 &  0.796670& 15.96500 & 11223 & 155.2 & 2.2 &  0.15 & 0.08& 0.37& 0.01& 13.27 &0.01\\ 
8 & 1 & 1237652944786292742 &  0.811670& 16.14556 & 1050 & 13.2 & 1.3 &  0.15 & 0.08& 0.79& 0.01& 10.46 &0.01\\ 
10 & 1 & 1237669682261983347 &  0.835420&  8.61861 & 11941 & 165.4 & 2.1 &  0.24 & 0.13& 0.89& 0.01& 13.41 &0.01\\ 
11 & 1 & 1237680297818390670 &  0.839580& 22.10250 & 4445 & 62.8 & 2.4 &  0.17 & 0.09& 0.72& 0.01& 14.80 &0.01\\ 
12 & 1 & 1237663234988769589 &  0.835000& 29.79722 & 6980 & 95.0 & 2.4 &  0.19 & 0.10& 0.59& 0.01& 14.26 &0.01\\ 
13 & 1 & 1237663234452947057 &  0.871670& 27.35139 & 7749 & 105.9 & 2.2 &  0.18 & 0.09& 0.78& 0.01& 13.52 &0.01\\ 
14 & 1 & 1237679478544400465 &  0.896250& 23.20028 & 7247 & 98.6 & 2.2 &  0.31 & 0.16& 0.68& 0.01& 13.17 &0.01\\ 
15 & 1 & 1237678660887511183 &  0.936250&  4.29806 & 11566 & 160.1 & 2.1 &  0.08 & 0.04& 0.36& 0.01& 14.23 &0.01\\ 
16 & 1 & 1237669680651436052 &  0.953750&  7.47861 & 5243 & 74.1 & 2.4 &  0.21 & 0.11& 1.00& 0.01& 11.85 &0.01\\ 
17 & 1 & 1237652943712551135 &  0.930000& 15.21805 & 876 & 20.3 & 4.3 &  0.16 & 0.08& 0.65& 0.01& 15.11 &0.01\\ 
19 & 1 & 1237680247351672949 &  0.995830& 20.75194 & 2308 & 33.2 & 4.2 &  0.19 & 0.10& 0.30& 0.01& 12.16 &0.01\\ 
21 & 1 & 1237669680651436249 &  1.037080&  7.37889 & 6198 & 87.7 & 2.3 &  0.21 & 0.11& 0.89& 0.01& 13.51 &0.01\\ 
22 & 1 & 1237679319089807445 &  1.020830& 10.29417 & 7749 & 105.6 & 2.2 &  0.33 & 0.17& 0.62& 0.01& 14.06 &0.01\\ 
23 & 1 & 1237679319626678329 &  1.054170& 10.79028 & 7974 & 108.8 & 2.2 &  0.28 & 0.15& 0.47& 0.01& 13.37 &0.01\\ 
24 & 1 & 1237680298355327071 &  1.061250& 22.58778 & 4446 & 48.1 & 9.1 &  0.23 & 0.12& 0.60& 0.01& 14.26 &0.01\\ 
25 & 1 & 1237678779015495755 &  1.104580&  6.17778 & 5072 & 78.7 & 15.2 &  0.23 & 0.12& 0.22& 0.01& 14.55 &0.01\\ 
26 & 1 & 1237663308527763474 &  1.102080& 31.47194 & 4955 & 42.1 & 8.3 &  0.19 & 0.10& 0.61& 0.01& 12.95 &0.01\\ 
27 & 1 & 1237678663035125847 &  1.121250&  5.84556 & 3113 & 44.5 & 9.2 &  0.17 & 0.09& 0.48& 0.01& 13.84 &0.01\\ 
30 & 1 & 1237680479804260595 &  1.136250& 33.55917 & 4766 & 68.7 & 4.2 &  0.17 & 0.09& 0.26& 0.01& 14.76 &0.01\\ 
31 & 1 & 1237679455999361146 &  1.217500& 17.19222 & 1034 & 12.7 & 4.3 &  0.10 & 0.05& 0.73& 0.01& 13.75 &0.01\\ 
34 & 1 & 1237669680114696248 &  1.275830&  6.92000 & 6139 & 105.7 & 21.8 &  0.17 & 0.09& 0.73& 0.01& 12.29 &0.01\\ 
\bottomrule 
\hline 
\end{tabular} 
\end{center} 
\end{table*}



Table 2 presents derived properties of the cross-listed objects in the ALFALFA-SDSS Catalog, including absolute magnitude, color, stellar mass, HI mass, and star formation rate (SFR). Stellar masses and SFR have been calculated using the methods described in sections \ref{sec:stellmass}, \ref{sec:IRstellmass}, and \ref{sec:IRsfr}. We also provide stellar mass and SFR from GSWLC-2, when available. Table 2 is organized as follows:

\begin{enumerate}
\item Column 1---AGC number (entry number in the AGC catalog)
\item Column 2---g-band internal extinction factor $\gamma_g$ in mag, as described in section 2.2
\item Column 3---i-band internal extinction factor $\gamma_i$ in mag, as described in section 2.2
\item Column 4---corrected absolute i-band magnitude in mag obtained using SDSS i-band cmodel magnitude corrected for galactic and internal extinction, as described in section 2.2
\item Column 5---uncertainty in corrected absolute i-band magnitude
\item Column 6---corrected g-i color obtained in mag using SDSS g- and i-bands model magnitude corrected for galactic and internal extinction, as described in section 2.2
\item Column 7---uncertainty in corrected g-i color 
\item Column 8---stellar mass from SDSS optical photometry in logarithmic solar units using Taylor method log M$_{\star, Taylor}$, as described in section 2.3
\item Column 9---uncertainty in log M$_{\star, Taylor}$
\item Column 10---stellar mass from infrared unWISE photometry in logarithmic solar units using McGaugh method log M$_{\star, McGaugh}$, as described in section 3.1
\item Column 11---uncertainty in log M$_{\star, McGaugh}$
\item Column 12---Stellar mass from GSWLC-2 in logarithmic solar units
\item Column 13---uncertainty in log M$_{\star, GSWLC-2}$
\item Column 14---Star Formation Rate from unWISE infrared photometry using flux at the frequency at 22 $\micron$ SFR$_{22}$ in logarithmic \msun yr$^{-1}$, as described in section 3.2
\item Column 15---uncertainty in SFR$_{22}$
\item Column 16---Corrected near-ultraviolet Star Formation Rate from GALEX NUV photometry SFR$_{NUV_{cor}}$ in logarithmic \msun yr$^{-1}$ for galaxies with NUV fluxes available in NASA-Sloan Atlas, as described in section 3.2
\item Column 17---uncertainty in SFR$_{NUV_{cor}}$
\item Column 18---Star Formation Rate from GSWLC-2 in logarithmic \msun yr$^{-1}$
\item Column 19---uncertainty in SFR$_{GSWLC-2}$
\item Column 20---HI mass in logarithmic solar units log M$_{HI}$ from \citet{haynes18a}
\item Column 21---uncertainty in log M$_{HI}$ from \citet{haynes18a}
\end{enumerate}

The full table is available online at the following website: http://egg.astro.cornell.edu/alfalfa/data/index.php.

\begin{sidewaystable*}
\begin{center}
\tablewidth{0.5\textwidth} 
\scriptsize 
\setlength\tabcolsep{1.0pt} 
\caption{Derived Properties of Cross-listed objects in the ALFALFA-SDSS Catalog.\label{tab:catalog2}  This table is available in its entirety online.}
\begin{tabular}{|c|c|c|c|c|c|c|c|c|c|c|c|c|c|c|c|c|c|c|c|c|c|c|}
\hline
\toprule
AGC & $\gamma_g$ &  $\gamma_i$  & M$_{icorr}$ &	$\rm \sigma_{M_{icorr}}$ &	(g-i)$_{corr}$	& $\sigma_{(g-i)_{corr}}$ & log M$_{\star}$ &	$\rm \sigma_{log M_{\star}}$  & log M$_{\star}$ &	$\rm \sigma_{log M_{\star}}$ & log M$_{\star}$& $\rm \sigma_{log M_{\star}}$ & logSFR$_{22}$ & $\rm \sigma_{log SFR_{22}}$  & logSFR$\rm _{NUVcor}$ & $\rm \sigma_{log SFR_{NUVcor}}$  & logSFR& $\rm \sigma_{logSFR}$ & M$_{HI}$ & $\rm \sigma_{M_{HI}}$  \\
&   & &  & &	& & Taylor & Taylor  & McGaugh & McGaugh &  GSWLC &GSWLC& & &  & & GSWLC & GSWLC &  &   \\
 & mag & mag & mag &  & mag & & $log(M_\odot)$ &  & $log(M_\odot)$ & & $log(M_\odot)$&  & $\rm log(M_\odot~yr^{-1})$ &  & $\rm log(M_\odot~yr^{-1})$  &  &  $log(M_\odot~yr^{-1})$&  & $log(M_\odot)$ &     \\
(1) & (2) & (3) & (4) & (5) & (6) & (7) & (8) & (9) & (10) & (11) & (12) & (13) & (14) & (15) & (16) & (17) & (18) & (19) &(20) &(21)  \\
\midrule
\hline
 1 & 0.73 & 0.39 & -19.62 & 0.07& 0.48  & 0.06 & 9.34 & 0.05 & 9.78& 0.02&-99.00 &-99.00 &-99.00 &-99.00 &-99.00 &-99.00&-99.00&-99.00&9.67&0.05  \\ 
 3 & 1.40 & 0.77 & -22.39 & 0.11& 1.06  & 0.14 & 10.85 & 0.11 & 10.89& 0.02&-99.00 &-99.00 &-0.40 &0.07 &-0.49 &0.03&-99.00&-99.00&10.28&0.05  \\ 
 4 & 1.25 & 0.68 & -21.66 & 0.07& 0.83  & 0.09 & 10.39 & 0.07 & 10.62& 0.02&-99.00 &-99.00 &0.12 &0.03 &0.01 &0.02&-99.00&-99.00&9.74&0.06  \\ 
 6 & 1.21 & 0.64 & -21.38 & 0.07& 0.77  & 0.07 & 10.24 & 0.05 & 11.23& 0.02&-99.00 &-99.00 &1.50 &0.02 &1.38 &0.03&-99.00&-99.00&9.29&0.07  \\ 
 7 & 1.59 & 0.86 & -23.14 & 0.13& 1.00  & 0.19 & 11.10 & 0.14 & 11.21& 0.01&11.39 &0.01 &-0.06 &0.05 &-0.16 &0.04&0.01&0.12&10.36&0.05  \\ 
 8 & 0.69 & 0.48 & -20.26 & 0.22& 1.64  & 0.05 & 10.41 & 0.09 & 9.32& 0.09&-99.00 &-99.00 &-1.97 &0.09 &-2.07 &0.09&-99.00&-99.00&8.96&0.10  \\ 
 10 & 1.70 & 0.87 & -22.86 & 0.04& 1.11  & 0.04 & 11.07 & 0.03 & 11.22& 0.01&-99.00 &-99.00 &0.44 &0.03 &0.33 &0.03&-99.00&-99.00&10.64&0.05  \\ 
 11 & 0.55 & 0.34 & -19.33 & 0.10& 0.65  & 0.07 & 9.33 & 0.06 & 9.58& 0.03&-99.00 &-99.00 &-2.18 &0.66 &-1.87 &0.04&-99.00&-99.00&9.25&0.06  \\ 
 12 & 1.01 & 0.56 & -20.85 & 0.09& 0.77  & 0.10 & 10.03 & 0.08 & 10.29& 0.02&-99.00 &-99.00 &-0.15 &0.03 &-99.00 &0.02&-99.00&-99.00&9.79&0.05  \\ 
 13 & 1.27 & 0.70 & -21.77 & 0.06& 1.05  & 0.06 & 10.59 & 0.05 & 10.79& 0.02&-99.00 &-99.00 &0.12 &0.02 &0.01 &0.03&-99.00&-99.00&9.24&0.08  \\ 
 14 & 1.41 & 0.74 & -22.08 & 0.08& 0.94  & 0.08 & 10.64 & 0.06 & 10.86& 0.02&-99.00 &-99.00 &0.57 &0.03 &0.45 &0.03&-99.00&-99.00&10.18&0.05  \\ 
 15 & 1.29 & 0.72 & -22.15 & 0.14& 1.12  & 0.19 & 10.79 & 0.14 & 10.90& 0.01&-99.00 &-99.00 &0.02 &0.05 &-0.09 &0.04&-99.00&-99.00&10.27&0.05  \\ 
 16 & 1.59 & 0.84 & -22.61 & 0.07& 1.19  & 0.03 & 11.02 & 0.04 & 10.16& 0.03&-99.00 &-99.00 &-0.75 &0.04 &-0.80 &0.03&-99.00&-99.00&10.18&0.05  \\ 
 17 & 0.00 & 0.00 & -16.51 & 0.46& 0.44  & 0.09 & 8.06 & 0.20 & 8.61& 0.18&-99.00 &-99.00 &-3.24 &1.08 &-2.78 &0.18&-99.00&-99.00&8.81&0.19  \\ 
 19 & 0.85 & 0.53 & -20.82 & 0.32& 1.04  & 0.22 & 10.20 & 0.20 & 10.81& 0.11&-99.00 &-99.00 &0.03 &0.11 &-99.00 &0.11&-99.00&-99.00&9.59&0.12  \\ 
 21 & 1.21 & 0.65 & -21.34 & 0.06& 0.91  & 0.04 & 10.32 & 0.04 & 10.49& 0.02&-99.00 &-99.00 &-0.07 &0.03 &-0.18 &0.03&-99.00&-99.00&9.95&0.05  \\ 
 22 & 1.10 & 0.63 & -21.36 & 0.09& 1.07  & 0.10 & 10.45 & 0.07 & 10.51& 0.02&-99.00 &-99.00 &-1.13 &0.25 &-99.00 &-99.00&-99.00&-99.00&9.88&0.05  \\ 
 23 & 1.37 & 0.74 & -22.20 & 0.11& 0.99  & 0.14 & 10.73 & 0.11 & 10.91& 0.02&-99.00 &-99.00 &0.26 &0.02 &0.14 &0.04&-99.00&-99.00&9.77&0.05  \\ 
 24 & 0.54 & 0.34 & -19.34 & 0.42& 0.58  & 0.10 & 9.29 & 0.18 & 9.57& 0.16&-99.00 &-99.00 &-1.53 &0.28 &-1.53 &0.17&-99.00&-99.00&9.45&0.17  \\ 
 25 & 0.75 & 0.46 & -20.35 & 0.46& 0.78  & 0.28 & 9.84 & 0.27 & 10.16& 0.17&-99.00 &-99.00 &-0.28 &0.17 &-0.39 &0.17&-99.00&-99.00&9.72&0.17  \\ 
 26 & 0.76 & 0.49 & -20.37 & 0.43& 0.89  & 0.10 & 9.92 & 0.19 & 10.16& 0.17&-99.00 &-99.00 &0.15 &0.17 &0.03 &0.17&-99.00&-99.00&9.63&0.18  \\ 
 27 & 0.64 & 0.37 & -19.61 & 0.46& 0.85  & 0.14 & 9.59 & 0.21 & 9.69& 0.18&-99.00 &-99.00 &-0.89 &0.18 &-0.98 &0.18&-99.00&-99.00&9.80&0.18  \\ 
 30 & 0.47 & 0.38 & -19.73 & 0.22& 0.83  & 0.25 & 9.62 & 0.20 & 9.62& 0.05&-99.00 &-99.00 &-1.00 &0.09 &-99.00 &-99.00&-99.00&-99.00&9.46&0.07  \\ 
 31 & 0.00 & 0.00 & -16.82 & 0.74& 0.77  & 0.07 & 8.41 & 0.30 & 8.42& 0.29&-99.00 &-99.00 &-2.66 &0.33 &-2.65 &0.29&-99.00&-99.00&8.01&0.30  \\ 
 34 & 1.62 & 0.89 & -23.04 & 0.45& 1.18  & 0.07 & 11.19 & 0.19 & 11.21& 0.18&-99.00 &-99.00 &-0.31 &0.18 &-0.41 &0.18&-99.00&-99.00&9.69&0.19  \\ 
\bottomrule
\hline
\end{tabular}
\end{center} 
\end{sidewaystable*} 


\section{Comparison with other Catalogs}\label{sec:comparison}

As a blind radio survey, the ALFALFA population is selected for HI gas content, with an additional bias toward galaxies with narrow HI line widths \citep{giovanelli05a}. In this section, we place the ALFALFA-SDSS catalog in the context of three other galaxy catalogs that include stellar masses: the NSA, S4G, and GSWLC-2 catalogs.  Our goal is to emphasize the overall differences between the ALFALFA-SDSS galaxy population and those from catalogs with selection effects that are related to different physical properties. 

\begin{figure}
    \centering
    \includegraphics[width=0.45\textwidth]{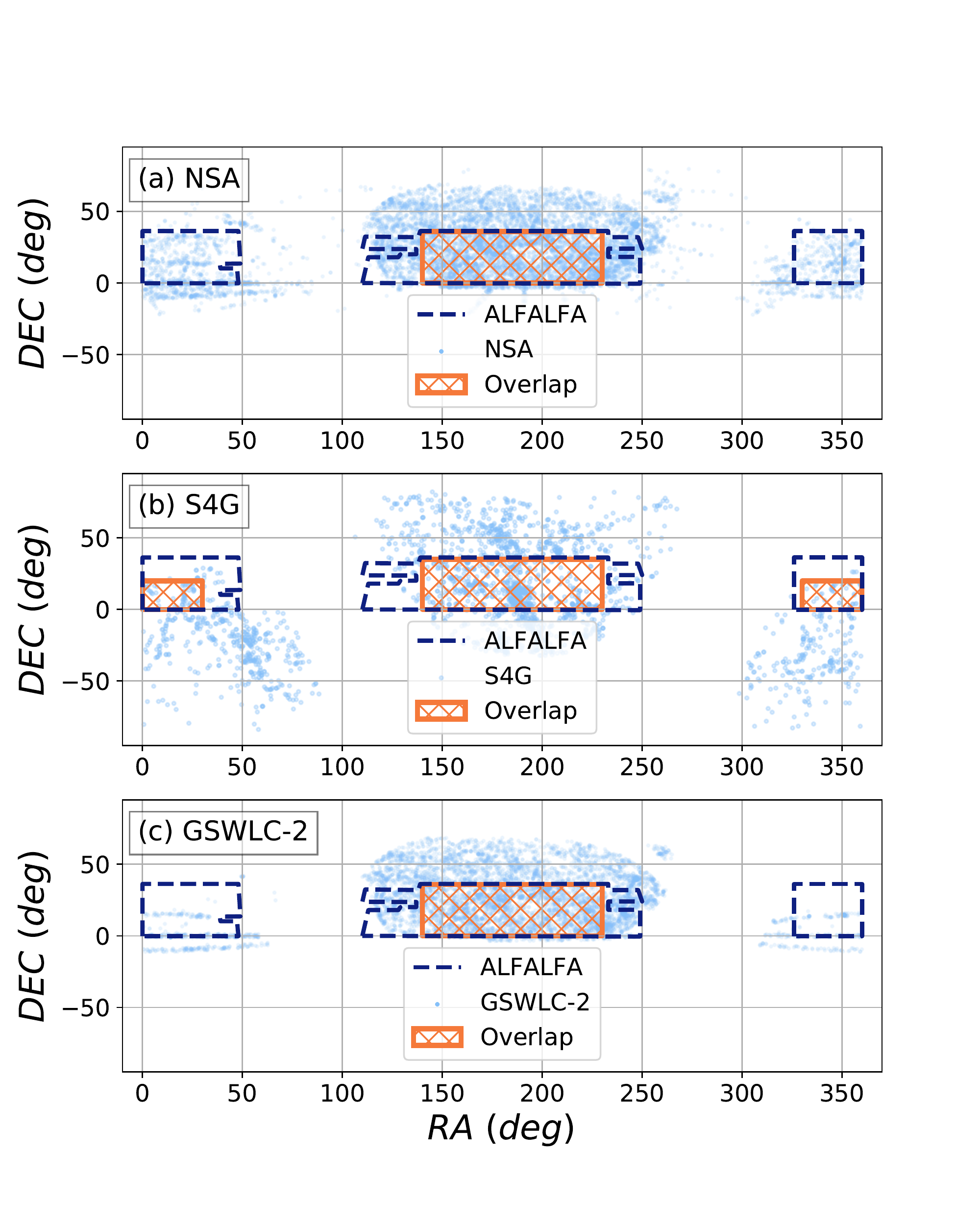}
    \caption{Dec. versus R.A. of galaxies in (top) NSA,  (center) GSWLC-2, and (bottom) S4G with cz $< 15,000$ \kms \ (NSA and GSWLC-2) and cz $< 3,000$ \kms (S4G) shown in light blue symbols. The dark blue dashed line traces the area covered by ALFALFA. The orange hatched areas show the overlap regions used to compare ALFALFA with each survey. See text for exact RA and dec coordinates of the regions used for comparison.} 
   \label{fig:RA-dec-a100-gswlc}
\end{figure}

For the purpose of comparing the ALFALFA-SDSS galaxy population with each of the other three catalogs, we limit the galaxies to a volume where the surveys overlap in order to account for differences in local environment (e.g. clusters versus voids) and differences in limiting distance. 
Figure \ref{fig:RA-dec-a100-gswlc} outlines the overlap volumes between ALFALFA and the comparison catalogs using orange hatched rectangles. The exact right ascension (RA) and declination (DEC) ranges of each region are provided in the following subsections. The galaxy populations of each catalog are displayed with light blue symbols in each panel. The dark blue dashed lines trace the area in the sky covered by ALFALFA.

Two of the comparison catalogs, the NSA and the S4G, do not include a corresponding SDSS object ID.  For these we find cross-identifications by searching for matches where the position difference on the sky is less than 15\arcsec \ and the radial velocity difference is less than 300~\kms. Note that this matching process is different from the more involved process described in Section \ref{sec:sdsscounterparts} used to create the ALFALFA-SDSS catalog itself. For the third comparison catalog, GSWLC-2, we find cross-identifications by matching with the SDSS object ID.
Table 3 summarizes the population statistics by catalog, first for each catalog separately, and then for the overlap volumes.

\subsection{Comparison with the NSA}

The NASA-Sloan Atlas version $v1\_0\_1$ \citep[NSA; ][]{blanton11a} includes images, photometric parameters, and spectroscopic parameters based on SDSS and GALEX data for 641,409 nearby galaxies to a redshift of $z=0.15$. Calculations are optimized for galaxies out to this redshift, and are designed to be an improvement over those in the SDSS pipeline. Future versions of the NSA are expected to incorporate information from the Two-Micron All Sky Survey (2MASS) and WISE. 
Absolute magnitudes in the NSA include k-corrections based on photometry using the {\it kcorrect} package \citep{blanton07a}. The {\it kcorrect} package estimates metallicities, star formation rates, stellar masses, and mass-to-light ratios as it determines the best k-correction; these estimates are included in the NSA as well. 

\begin{deluxetable}{lcc}
\tabletypesize{\footnotesize}
\tablewidth{0.5\textwidth}
\tablecaption{Population Statistics by Catalog \label{tab:CatalogStats}}
\tablecolumns{3}
\tablehead{
\colhead{Catalog} &
\colhead{Total Number} &
\colhead{Matched with ALFALFA}
}
\startdata
\sidehead{Full catalogs}  
ALFALFA  & 31501  & 31501 \\
NSA   & 641409 &  22860\\
S4G   & 2352   & 736\\
GSWLC-2 & 640659 &  13368\\
\sidehead{Overlap comparison volumes}  
NSA : ALFALFA    & 42999 : 15467 & 13547\\
S4G : ALFALFA    & 607 : 16811    & 541 \\
GSWLC-2 : ALFALFA  & 34628 : 15467 & 10425\\
 \enddata
\tablecomments{For each overlap comparison volume, we list the total number of galaxies in each catalog within this volume and the number of ALFALFA galaxies within this volume, separated by a colon (column two) as well as the number of matched galaxies (column three).  The limits of these volumes in terms of RA, Dec, and redshift are described in the text.}
%
\end{deluxetable}


Based on a match of the entire catalog, we find that 22,860 NSA galaxies are also in ALFALFA. (The NSA $v1\_0\_1$ has a parameter for membership in the ALFALFA catalog, but this was based on an early, incomplete version of the catalog; there are many more matches now.)

\begin{figure*}[htp]
    \centering
    \includegraphics[width=0.48\textwidth]{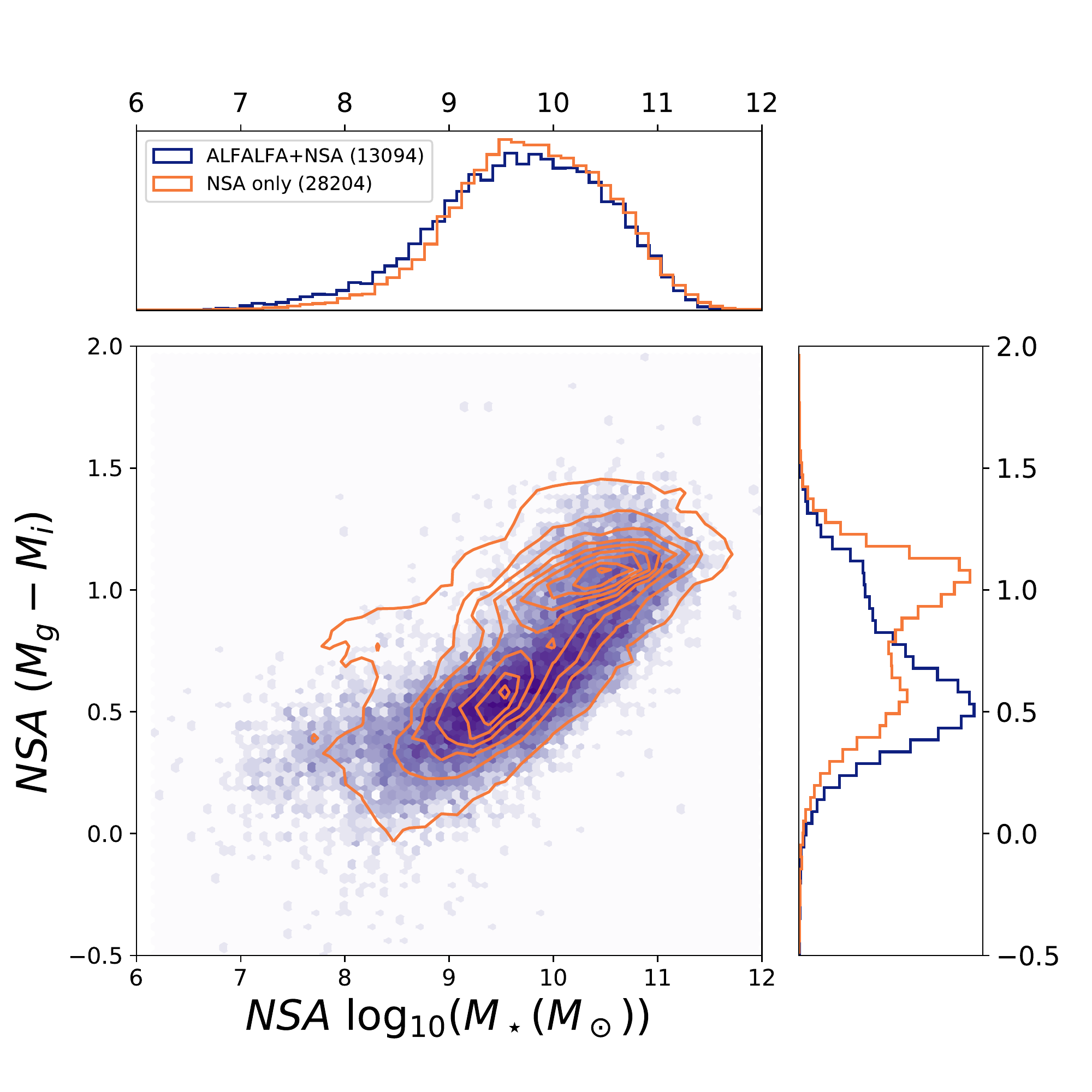}
    \includegraphics[width=0.48\textwidth]{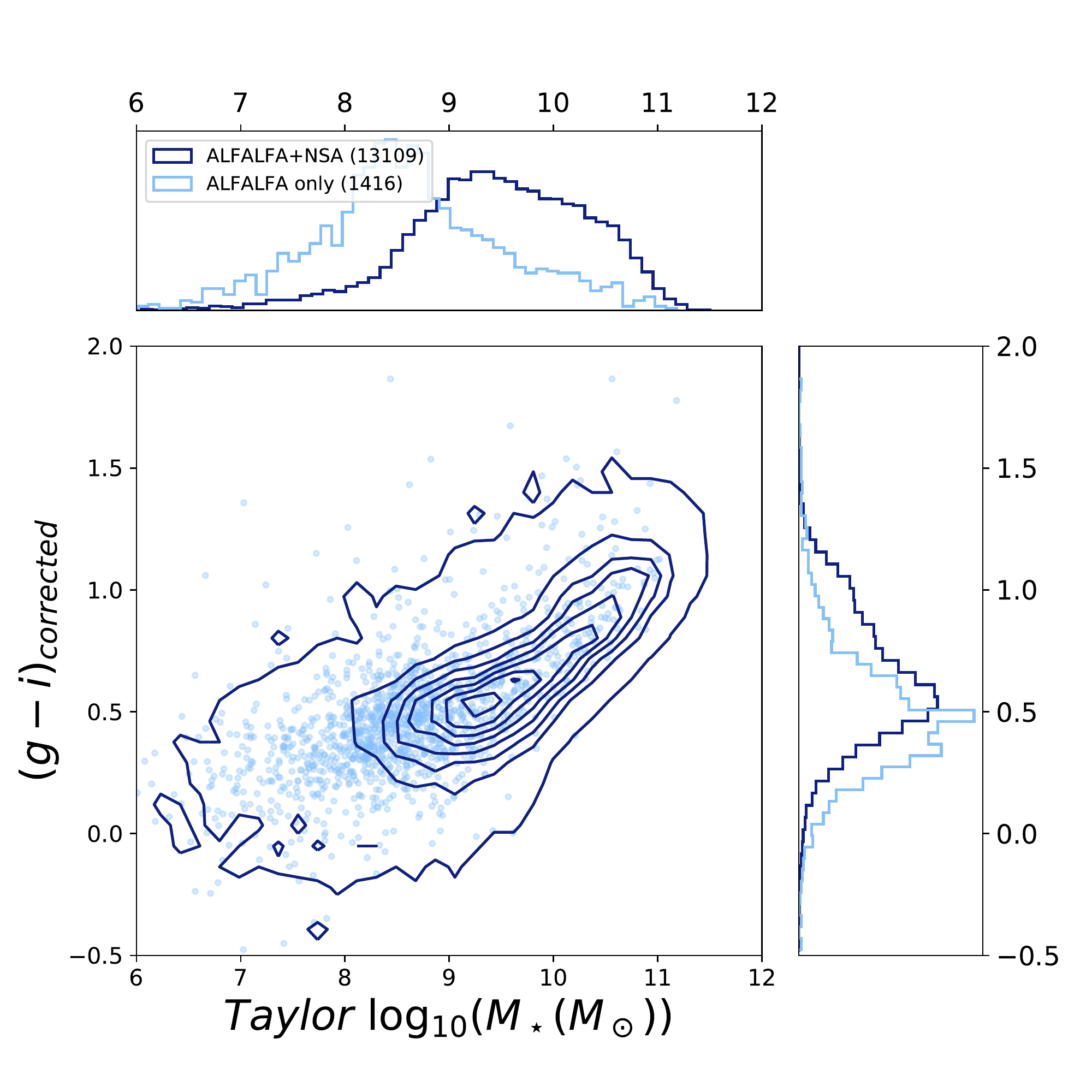}
    \caption{(Left) $M_g-M_i$ color vs. stellar mass (both from the NSA) for galaxies that lie in a volume common to both the ALFALFA and NSA surveys. Galaxies that are detected by both surveys are shown in blue, and the galaxies in the NSA but not in ALFALFA are shown as orange contours.  The ALFALFA sample is dominated by blue galaxies whereas the NSA is dominated by red galaxies.  (Right) Complementary comparison, showing $g-i$ corrected color vs. stellar mass calculated following \cite{taylor11a} for galaxies in common to both surveys (blue contours) and galaxies in ALFALFA but not in the NSA (light blue).  When compared to the NSA galaxies, ALFALFA galaxies are bluer and include more lower mass galaxies.} 
    \label{fig:nsa}
\end{figure*}

To compare the galaxy populations in ALFALFA and the NSA, we first select galaxies that lie within a volume that is common to both surveys.  Specifically, we use the following selection criteria: $z<0.05$, and 
$$140\degrees <  R.A. < 230\degrees,$$
$$0\degrees < Dec. < 35\degrees.$$
We compare color versus stellar mass in Figure \ref{fig:nsa}.  In the left panels, we show the distribution of galaxies that are common to both surveys (blue symbols) and galaxies that are in the NSA but not in ALFALFA (orange contours).  Note that in the left panels we show the $M_g-M_i$ color (corrected for galactic extinction but not internal extinction) and stellar mass that come from the NSA catalog ($SERSIC\_ABSMAG$, $EXTINCTION$ and $MASS$), because not all of the galaxies are in ALFALFA. The left plot and histograms of $g-i$ color show that the NSA galaxy population is dramatically different from the ALFALFA galaxy population in that it includes the red sequence as well as the blue cloud. In the right panels, we show the complementary comparison, namely galaxies that are in ALFALFA but not in the NSA (light blue symbols), and we again compare to the population that is in common (blue contours). Note that in the right-hand panels we show our values of $g-i$ (section \ref{sec:optphot}) and $M_\star$ (section \ref{sec:stellmass}), because not all of the galaxies are in the NSA. The ALFALFA galaxies that are not in the NSA (lighter blue symbols) are lower-mass and bluer than those that are in the NSA (blue symbols). 

\subsection{Comparison with the S4G }

\begin{figure*}[htp]
    \centering
    \includegraphics[width=0.48\textwidth]{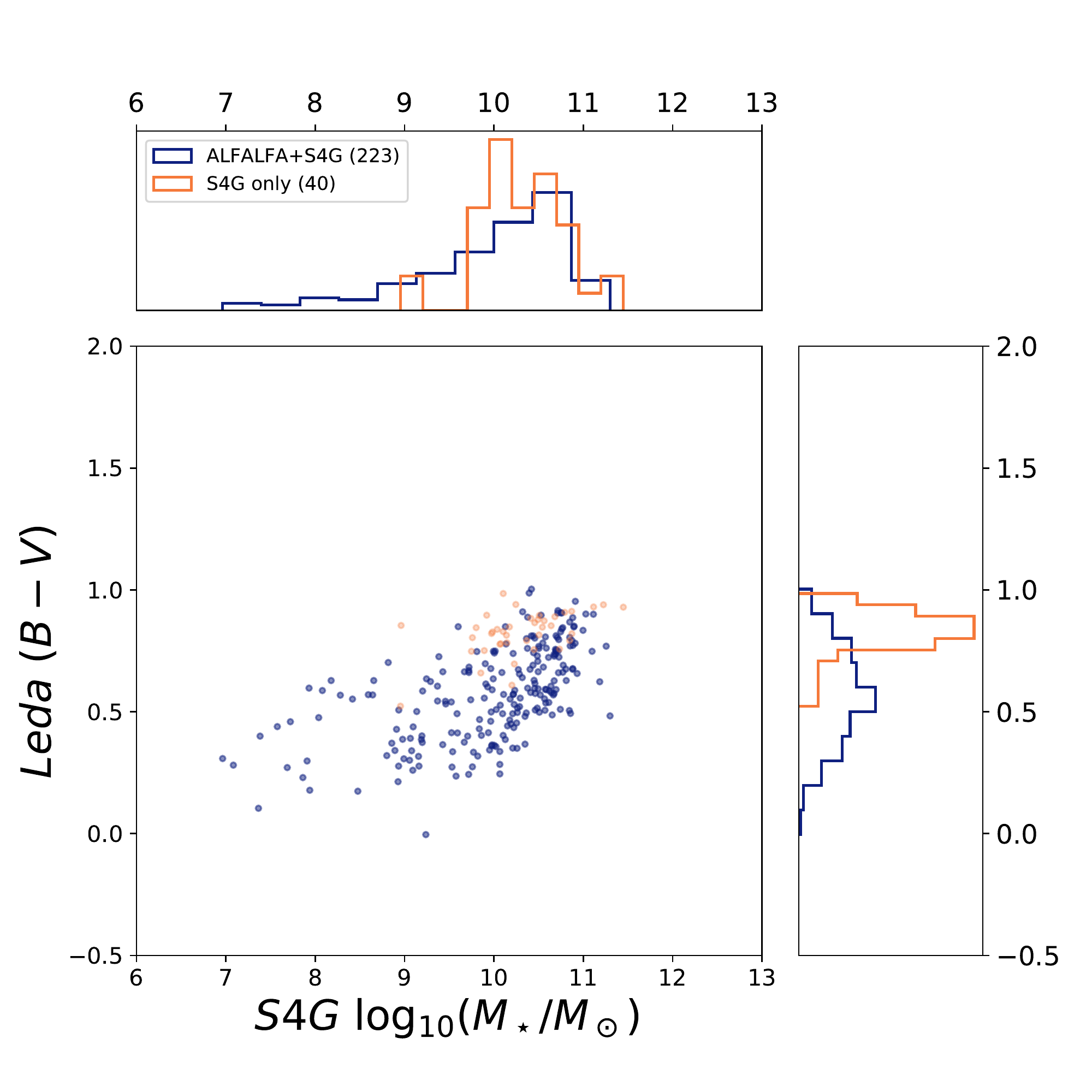}
    \includegraphics[width=0.48\textwidth]{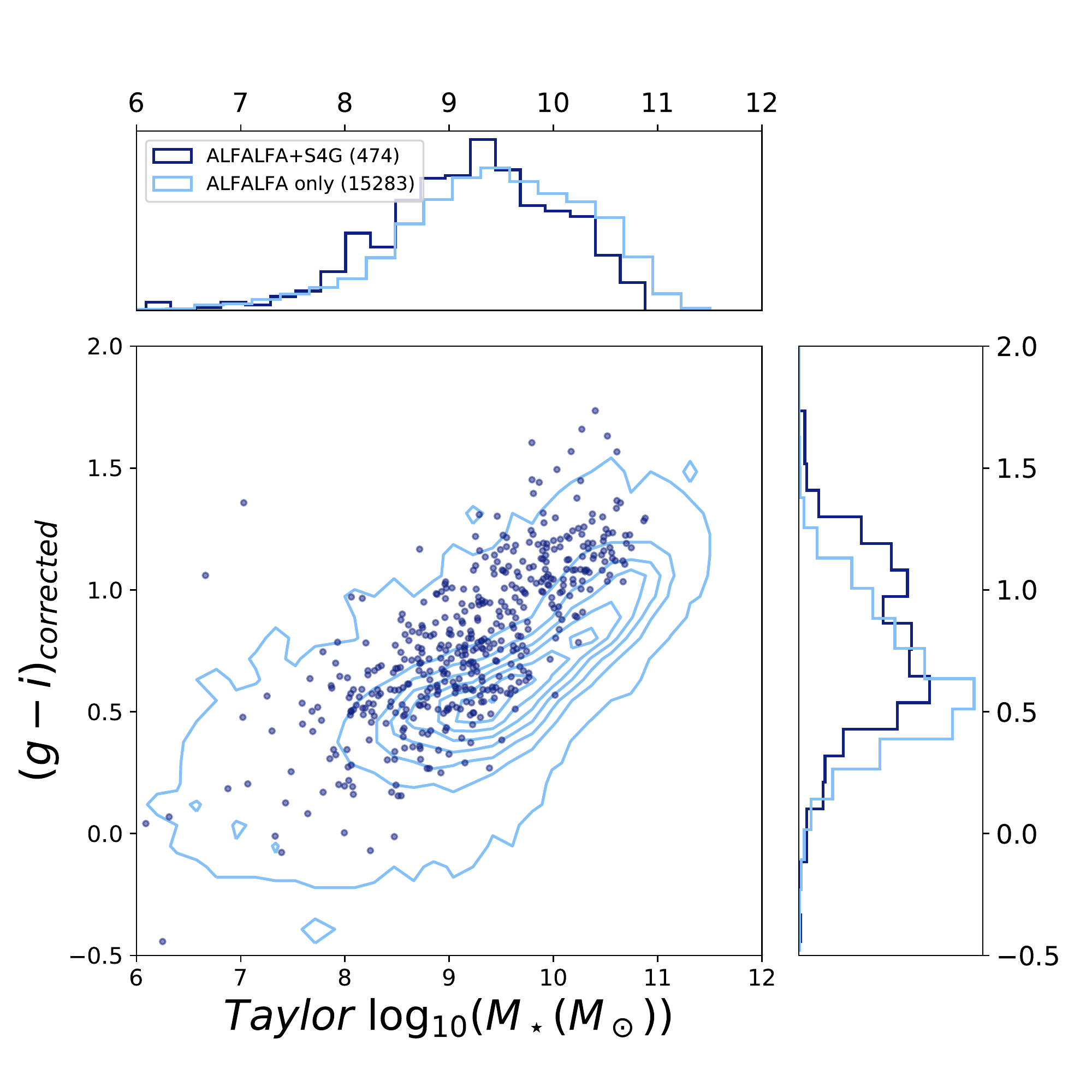}
    \caption{(Left) $B-V$ color vs. stellar mass (both from the S4G catalog) for galaxies that lie in a volume common to both the ALFALFA and S4G \citep{sheth10} surveys.  Galaxies that are detected by both surveys are shown in blue, and galaxies in the S4G but not in ALFALFA are shown in orange. (Right) Complementary comparison, showing $g-i$ corrected color vs. stellar mass calculated following \cite{taylor11a} for galaxies in common to both surveys (blue) and galaxies in ALFALFA but not in the S4G (light blue contours). When compared to the S4G galaxies, ALFALFA galaxies are bluer and have similar masses.} 
    \label{fig:s4g}
\end{figure*}

The Spitzer Survey of Stellar Structure in Galaxies (S4G) is designed to provide a large baseline sample for the distribution of stellar mass within galaxies out to about 40 Mpc. It contains infrared images and photometric parameters for 2352 galaxies extending down to stellar masses $\sim 10^{7}$ \msun~  that have been mapped using the IRAC 3.6 and 4.5 micron channels \citep{sheth10, munos2013a, querejeta15a}.  It covers a much larger area of the sky than ALFALFA, but out to a smaller redshift of $z=0.01$.  
The overlap volume with ALFALFA used to compare galaxy populations is shown the middle panel of Figure \ref{fig:RA-dec-a100-gswlc}.  In the north galactic hemisphere, the overlap region is defined as $z<0.01$ and: \\
$$138\degrees <  R.A. < 232\degrees,$$ 
$$0\degrees < Dec. < 35\degrees.$$
We add an additional region in the southern galactic hemisphere of $z<0.01$ and : 
$$0\degree < R.A. < 30\degrees \  {\rm or}  \ 330 < R.A. < 360\degrees,$$ 
$$0\degrees < Dec. < 20\degrees.$$
The matching statistics are reported in Table \ref{tab:CatalogStats}, and the stellar masses and colors are shown in Figure \ref{fig:s4g}. While most of the S4G galaxies are small and blue relative to large optical surveys, ALFALFA galaxies are still bluer than the S4G population.  The samples have similar masses, although the resulting mass distributions are somewhat sensitive to the exact choice of overlap region. We investigated using slightly different criteria for the overlap region, including the same cut in redshift but  different boundaries for position on the sky. In each case, the ALFALFA galaxies were bluer. However, in some cases, the ALFALFA galaxies had, on average, slightly lower masses, and in other cases they had slightly higher masses.

\subsection{Comparison with the GSWLC}

The GALEX-SDSS-WISE Legacy Catalog \citep[GSWLC, ][]{salim16a, salim18a} contains galaxies within the GALEX footprint, whether or not they were detected in the UV. The catalog includes physical properties (e.g., stellar mass, SFR) of about 650,000 galaxies with SDSS redshifts below 0.3. There are two versions of the catalog: the GSWLC-1 and the  GSWLC-2. Both versions contain the same sources (exactly the same number in the same order) and they both use identical photometry. GSWLC-2 has more accurate SFRs from joint UV+optical+mid-IR SED fitting while GSWLC-1 contains separate star formation rates (SFRs) from the UV+optical SED fitting. For our comparison with ALFALFA we use GSWLC-2. The redshifts reported in the GSWLC-2 catalog are from SDSS.

The overlap volume employed to compare galaxy populations in GSWLC-2 and ALFALFA is the same as that for the NSA: $z<0.05$, and 
$$140\degrees <  R.A. < 230\degrees,$$ 
$$0\degrees < Dec. < 35\degrees.$$ 
This overlap region is outlined with an orange hatched rectangle in the bottom panel of Figure \ref{fig:RA-dec-a100-gswlc}.

\begin{figure*}[htp]
    \centering
    \includegraphics[width=0.48\textwidth]{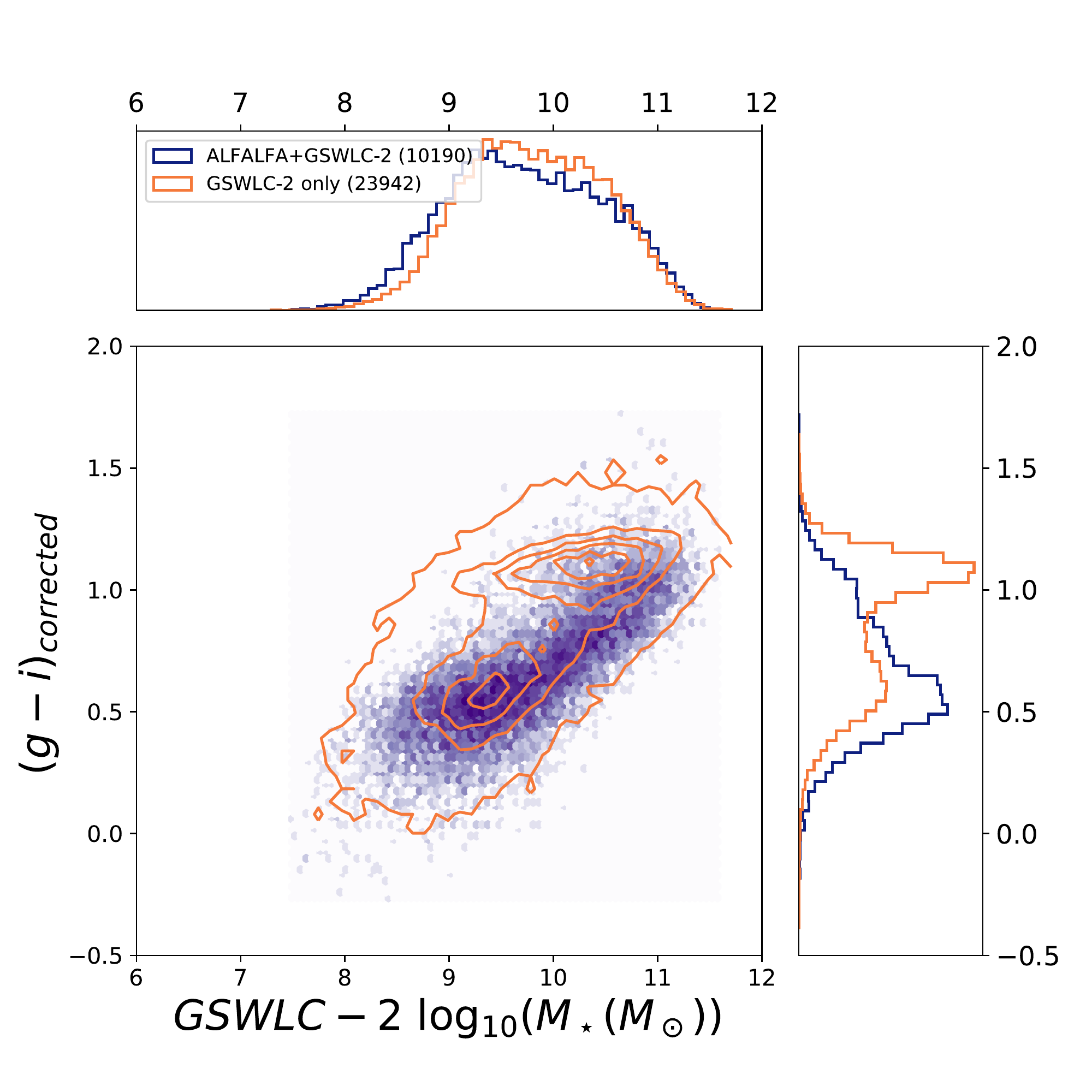}
    \includegraphics[width=0.48\textwidth]{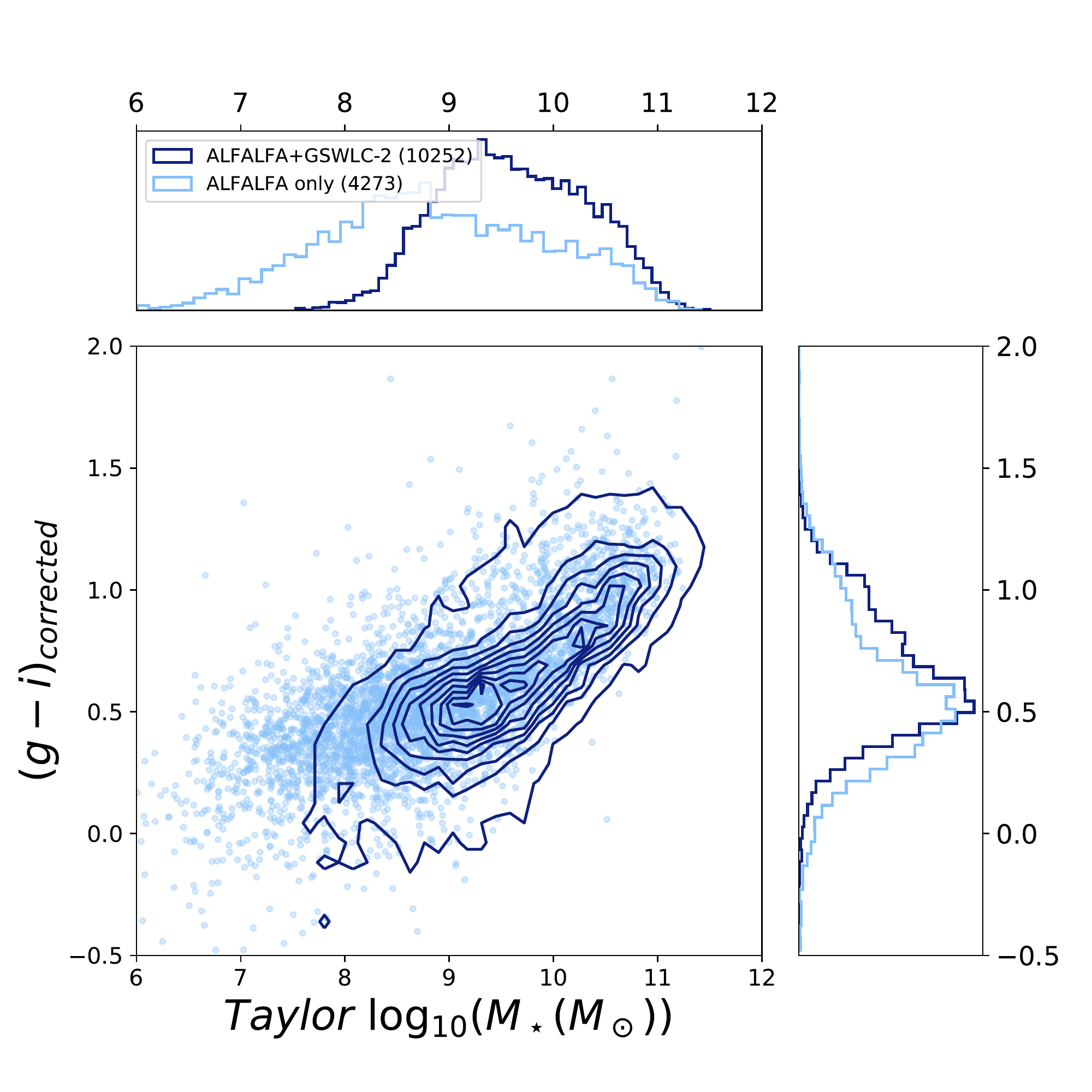}
    \caption{(Left) $g-i$ color vs stellar mass from the GSWLC-2 catalog for galaxies that lie in a volume common to both the ALFALFA and GSWLC-2 \citep{salim16a} surveys. Galaxies that are detected by both surveys are shown in blue, and the galaxies in the GSWLC-2 but not in ALFALFA are shown as orange contours. The ALFALFA sample is dominated by blue galaxies whereas the GSWLC-2 is dominated by red galaxies.  (Right) Complementary comparison, showing $g-i$ corrected color vs. stellar mass calculated following \cite{taylor11a} for galaxies in common to both surveys (blue contours) and galaxies in ALFALFA but not in the GSWLC-2 (light blue) as a function of stellar mass determined using the \citet{taylor11a} method. When compared to the GSWLC-2 galaxies, ALFALFA includes more lower mass (and slightly bluer) galaxies. }
    \label{fig:gswlc}
\end{figure*}

The matching statistics are reported in Table \ref{tab:CatalogStats},
and the colors and stellar masses are shown in Figure \ref{fig:gswlc}. For colors we use SDSS $g-i$ model magnitudes for both ALFALFA and GSWLC-2 galaxies, corrected for both galactic and internal extinction as described in section 2.2. The left panel uses stellar masses from the GSWLC-2 catalog while the right panel uses stellar mass values calculated using the \citet{taylor11a} method, as explained in section \ref{sec:stellmass}. The left panel compares galaxies found in both surveys (blue symbols) to galaxies found in GSWLC-2 but not in ALFALFA (orange contours). The right panel compares galaxies found in both surveys (blue contours) to galaxies found in ALFALFA but not in GSWLC-2 (light blue symbols).
Comparing the left and right panels of Figure \ref{fig:gswlc}, we see that galaxies in GSWLC-2 tend to be redder and more massive compared to the ALFALFA population.

\begin{figure*}[htp]
    \centering
    \includegraphics[width=0.48\textwidth]{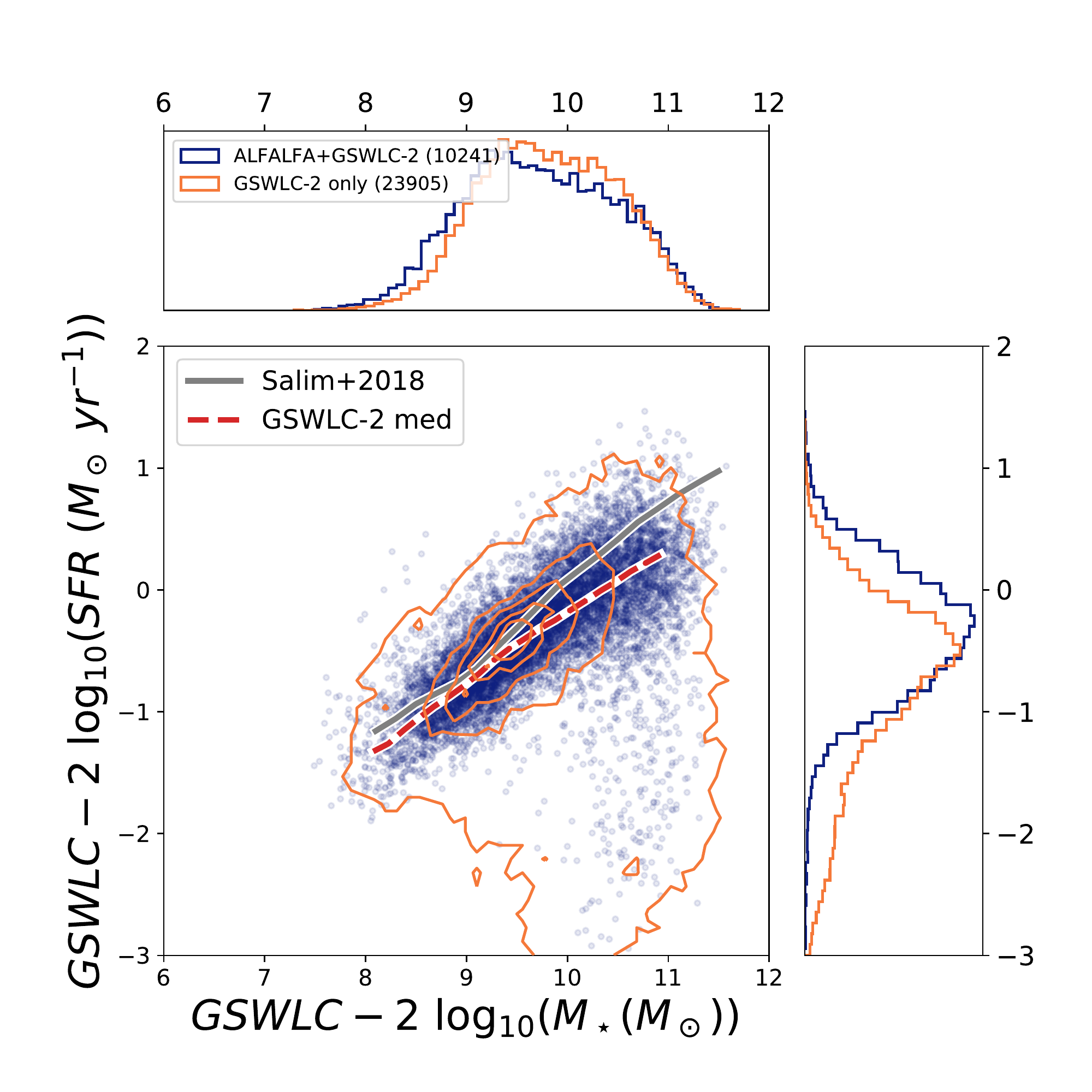}
    \includegraphics[width=0.48\textwidth]{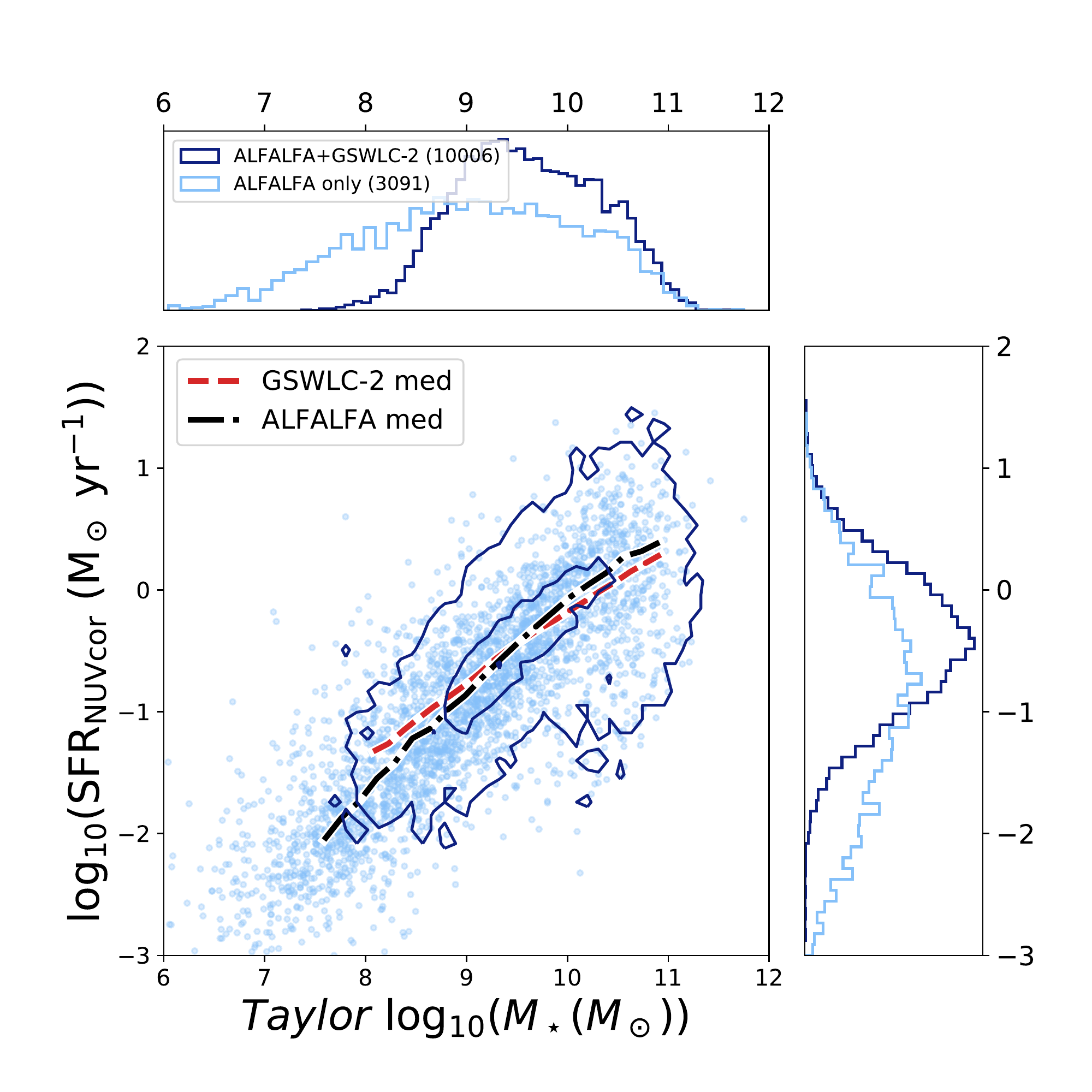}
    \caption{(Left) Star formation rate vs. stellar mass (both from the GSWLC-2 catalog) for galaxies in a common volume to both the ALFALFA and GSWLC-2 surveys. Galaxies that are detected in both surveys are shown in blue, and galaxies found in GSWLC-2 but not in ALFALFA are shown as orange contours. The red dashed line is the median for the star-forming GSWLC-2 galaxies in the overlap region and the gray line represents the fit to the full GSWLC-2 sample from \citet{salim18a}. The slopes are different for $\log_{10}(M_\star/M_\odot) > 9.5$, possibly due to the higher redshift cut for the GWSLC-2. (Right) Complementary comparison, showing the NUV-corrected star formation rate vs. stellar mass calculated following \cite{taylor11a} for galaxies in common to both surveys (blue contours) and galaxies found in ALFALFA but not in GSWLC-2 (light blue). The red dashed line is the median for the GSWLC-2 galaxies in the overlap region and the black dash-dotted line is the median star formation rate versus stellar mass for the ALFALFA galaxies. The star-forming main sequences for the GSWLC-2 and ALFALFA samples are similar, but with a slightly steeper slope for the ALFALFA galaxies that drops below the GSWLC-2 relation at the lowest masses (see text for discussion).  }
    \label{fig:sfr-mstar}
\end{figure*}

For the GSWLC-2, we also compare the galaxy populations in terms of star formation rates (SFRs) and specific star formation rates (sSFR).
In Figure \ref{fig:sfr-mstar}, we plot SFR versus stellar mass for the GSWLC-2 and ALFALFA galaxies. The left panel of Figure \ref{fig:sfr-mstar} compares galaxies that are found in both ALFALFA and GSWLC-2 (blue symbols) to those found in GSWLC-2 but not in ALFALFA (orange contours). The majority of the galaxies fall on the star-forming main sequence. 
The red dashed line in the left panel of Figure \ref{fig:sfr-mstar} shows the median SFR versus stellar mass for the star-forming GSWLC-2 galaxies in the overlap region, where we define star-forming as $\log_{10}(sSFR) > -11$ according to the criteria of \citet{salim18a}. For comparison, the gray line shows the main sequence derived from the full GSWLC-2 \citep{salim18a}. We find that the slopes are significantly different for  $\log_{10}(M_\star/M_\odot) > 9.5$. We attribute the offset between the two lines to evolutionary effects between the ALFALFA galaxies and the higher-redshift galaxies that are more typical of SDSS and thus the GSWLC-2.

In the right panel, we again show the galaxies that are in both the ALFALFA and GSWLC-2 samples (blue contours), but now compare to the galaxies that are in ALFALFA but not in GSWLC-2 (light blue symbols).  Here we use NUV-corrected SFRs and the \citet{taylor11a} stellar masses.
We show the median SFR vs.\ stellar mass for the ALFALFA star-forming galaxies ($\log_{10}(sSFR) > -11$) with the black dash-dotted line, and we again show the median SFR for the star-forming GSWLC-2 galaxies with the red dashed line for comparison.    
We see that the star-forming main sequences for the
GSWLC-2 and ALFALFA samples are similar.  However, there is a hint that the slope of the \aone \ relation is steeper and drops below the GSWLC-2 relation at the lowest stellar masses.

\begin{figure*}
    \centering
    \includegraphics[width=0.48\textwidth]{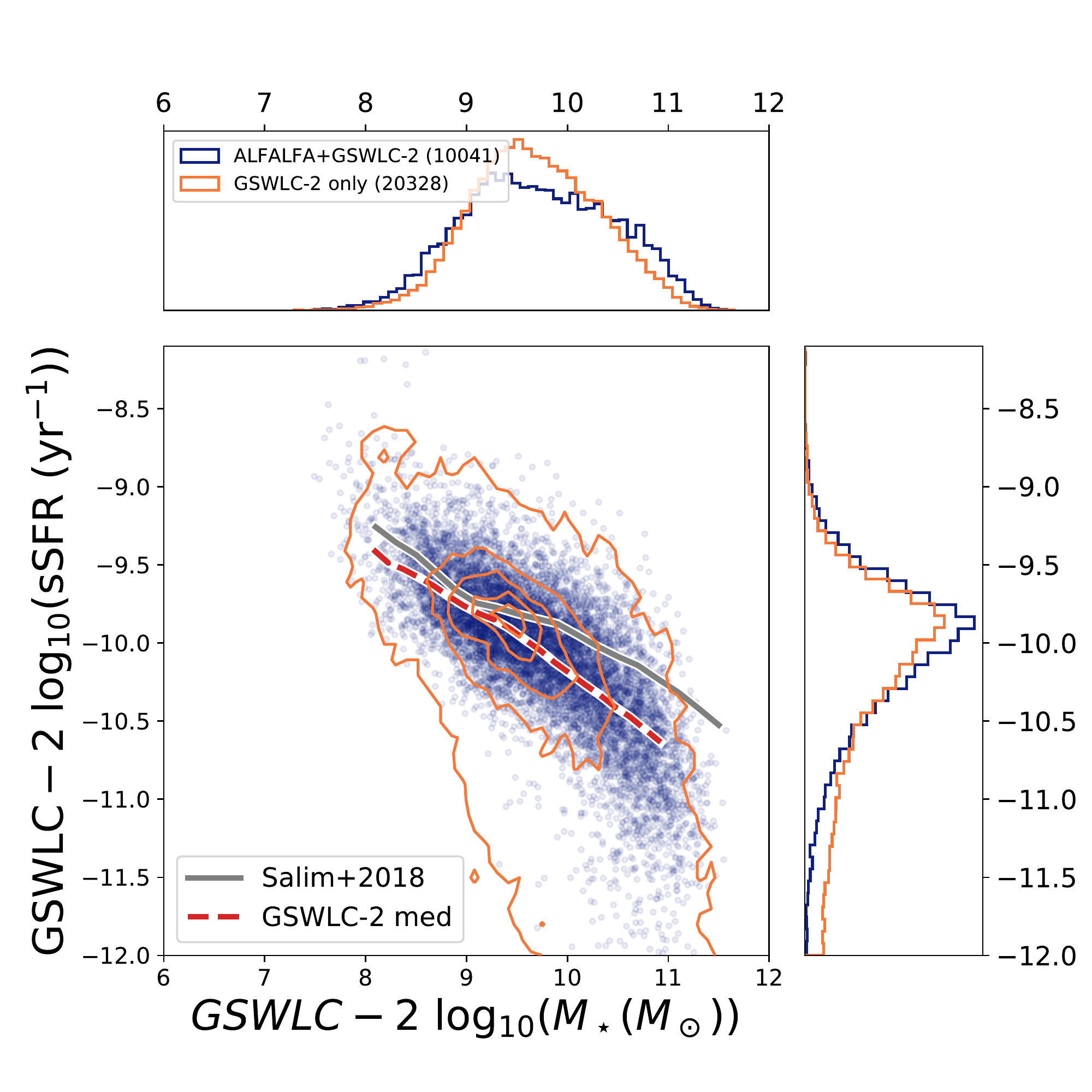}
    \includegraphics[width=0.48\textwidth]{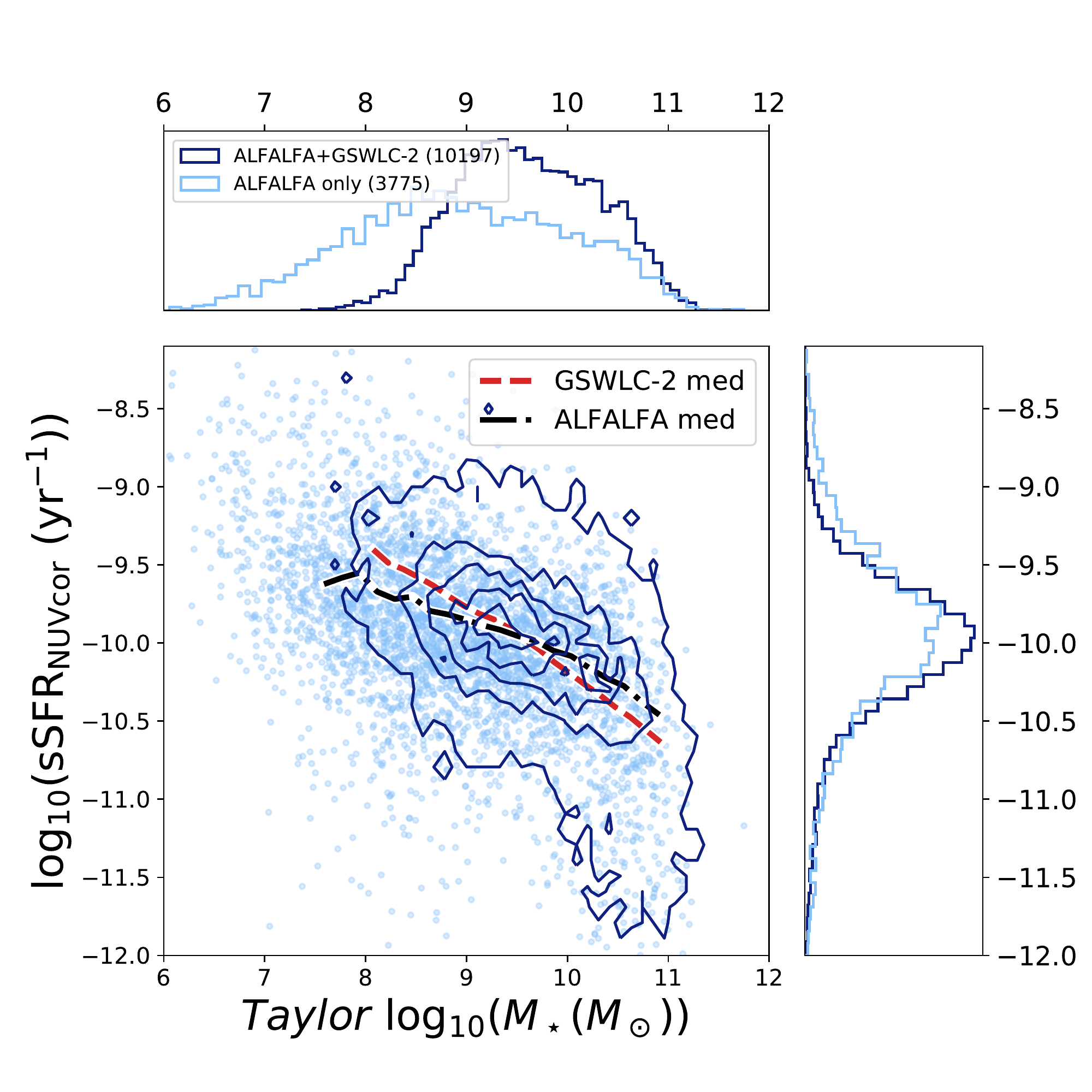}
    \caption{(Left) Specific star formation rate vs. stellar mass (both from the GSWLC-2 catalog) for galaxies in a common volume to both ALFALFA and GSWLC-2 surveys. Galaxies that are detected in both surveys are shown in blue, and galaxies in GSWLC-2 but not in ALFALFA are shown as orange contours. The red dashed line is the median for the GSWLC-2 galaxies in the overlap region and the gray line represents the fit to the full GSWLC-2 sample from \citet{salim18a}. The slopes are significantly different for  $\log_{10}(M_\star/M_\odot) > 9.5$, possibly because of the higher redshift cut of the GSWLC-2. (Right) Complementary comparison, showing the NUV-corrected specific star formation rate vs. stellar mass calculated following \cite{taylor11a} for galaxies in common to both surveys (blue contours) and galaxies in ALFALFA but not in GSWLC-2 (light blue). The red dashed line is the median for the GSWLC-2 galaxies in the overlap region and the black dash-dotted line is the median sSFR versus stellar mass for the ALFALFA galaxies. The median specific star formation rates of the ALFALFA galaxies are similar to those of the GSWLC-2 galaxies, but with a slightly shallower slope (see text for discussion).  }
    \label{fig:ssfr-mstar}
\end{figure*}

Similarly, in Figure \ref{fig:ssfr-mstar} we compare the sSFR versus stellar mass for GSWLC-2 and ALFALFA galaxies. The general trend is that as stellar mass increases, the sSFR decreases. 
In the left panel we show the median sSFR versus stellar mass for the GSWLC-2 galaxies in our overlap comparison sample with the red dashed line.  The gray line shows the relationship for the full GSWLC-2 \citep[Fig.~3 from][]{salim18a}.
The trends differ significantly at $\log_{10}(M_\star/M_\odot) > 9.5$, and we again attribute the offset to the lower redshift-cut that we apply to the overlap sample. 
The right panel of Figure \ref{fig:ssfr-mstar} shows the \aone \ galaxies.  The median sSFR of the ALFALFA galaxies (black dash-dotted line) is close to the GSWLC-2 galaxies, but again the \aone \ relation has a shallower slope. 

The difference in slopes between the ALFALFA and GSWLC-2 samples is intriguing.  We note that we have made no effort to correct for the incompleteness that we undoubtedly suffer in detecting galaxies with the lowest SFRs. However, such a correction would be expected to bring the ALFALFA main sequence even lower. Thus our results suggest a flatter relationship between sSFR and stellar mass than observed for the GSWLC-2 sample.
This difference might be due in part to the fact that ALFALFA is sensitive to lower-mass galaxies with high gas content but lower sSFRs \citep[e.g.][]{huang12b}.
However, a full analysis of completeness, both in terms of stellar mass and SFRs, is needed to constrain the behavior of the star-forming main sequence for dwarf galaxies.
Such an analysis is beyond the scope of this paper.

\section{Summary}\label{sec:Summary}
We present an HI-optical catalog of matches between the recently completed 100\% ALFALFA survey (ALFALFA) and SDSS. The ALFALFA-SDSS catalog contains 31,501 galaxies. We provide SDSS identifications for nearly all ALFALFA galaxies (29418 galaxies) including 12737 that are hard to identify because they don't have SDSS spectroscopy.  The SDSS identifications in the ALFALFA-SDSS catalog can be used as basis for further cross-matching with other surveys at other wavelengths. Most galaxies with SDSS photometry have uncertainties less than 0.05 mag in $g$ and $i$ (\textit{good photometry}) and are assigned a code of ``1'' in our catalog (28057 objects). Galaxies with uncertainties greater than 0.05 in $g$ and/or $i$ (\textit{bad photometry}) are assigned a code of ``2'' (1361 objects). Galaxies with no clear SDSS counterpart are assigned a code of ``0'' if they are outside SDSS footprint (1296 objects) and ``3'' otherwise (787 objects).
We present observed (Table \ref{tab:catalog1}) and derived properties (Table \ref{tab:catalog2}) for the entire \aone \ sample, including absolute magnitude, color, stellar mass, HI mass, and star formation rate. In addition, we include magnitude-dependent internal extinction estimates that differ significantly from previous work and are better suited to the low-mass galaxies that dominate the \aone \ sample.

We explore different methods to calculate stellar mass based on SDSS optical \citep{taylor11a} and unWISE infrared photometry \citep{mcgaugh15a, cluver14}. We find that the Taylor method using optical SDSS gives the best agreement with SED-derived stellar masses from the GSWLC-2.
We also explore different methods to calculate the star formation rate using unWISE infrared and/or NUV GALEX photometry \citep{kennicutt12a,hao11a, murphy11a}.  The corrected NUV SFR agrees most closely with the SFR estimates from the GSWLC-2 and should be used when available.

We place the ALFALFA-SDSS Catalog in the context of three other galaxy catalogs that include stellar mass (NSA, GSWLC-2, and S4G) and star formation rate (GSWLC-2). In this way we show how the ALFALFA-SDSS sample compares with other optically-selected catalogs.  We find that ALFALFA-SDSS galaxies are generally less massive and bluer. We further compare the \aone \ and GSWLC-2 samples in terms of the $SFR-M_\star$ and $sSFR-M_\star$ relations.  The median relationships are similar, but we find evidence for a population of low mass, low sSFR galaxies in \aone \ that are not included in the GSWLC-2.  In addition, the slope of the $sSFR-M_\star$ relation appears flatter for the \aone \ sample.  We emphasize that these comparisons are meant to show how the ALFALFA galaxy population differs overall from the populations in other catalogs.  An analysis of the true scaling relations among galaxy properties requires a more thorough analysis of the completeness of each survey, which we reserve for future work.  

\section{Acknowledgements} \label{sec:Acknowledgements}

We are grateful for the contributions of all members of the Undergraduate ALFALFA Team. The authors thank D. Lang and S. Salim for assistance with utilizing their catalogs. This work has been supported by NSF grants AST-1211005, AST-1637339 and AST-1637271. The ALFALFA team at Cornell has been supported by NSF grants AST-1107390 and AST-1714828 and by the Brinson Foundation. RAF gratefully acknowledges support from NSF grants AST-0847430 and AST-1716657.

From 2011-2018, the Arecibo Observatory was operated by SRI International under a cooperative agreement 
with the National Science Foundation (AST-1100968), and in alliance with Ana G. Méndez-Universidad Metropolitana, and the Universities Space Research Association.  Currently,
the Arecibo Observatory is a facility of the National Science Foundation operated under cooperative 
agreement (AST-1744119) by the University of Central Florida in alliance with Universidad Ana G. Méndez (UAGM) and Yang Enterprises (YEI), Inc.

Funding for the SDSS and SDSS-II has been provided by the Alfred P. Sloan Foundation, the participating institutions, the National Science Foundation, the US Department of Energy, the NASA, the Japanese Monbukagakusho, the Max Planck Society, and the Higher Education Funding Council for England. The SDSS Web site is http://www.sdss.org/. The SDSS is managed by the Astrophysical Research Consortium
 for the participating institutions. 
 
Funding for the NASA-Sloan Atlas has been provided by the NASA Astrophysics Data Analysis Program (08-ADP08-0072) and the NSF (AST-1211644)
 
Funding for SDSS-III has been provided by the Alfred P. Sloan Foundation, the Participating Institutions, the National Science Foundation, and the U.S. Department of Energy. The SDSS-III web site is http://www.sdss3.org.
SDSS-III is managed by the Astrophysical Research Consortium for the Participating Institutions of the SDSS-III Collaboration including the University of Arizona, the Brazilian Participation Group, Brookhaven National Laboratory, University of Cambridge, University of Florida, the French Participation Group, the German Participation Group, the Instituto de Astrofisica de Canarias, the Michigan State/Notre Dame/JINA Participation Group, Johns Hopkins University, Lawrence Berkeley National Laboratory, Max Planck Institute for Astrophysics, New Mexico State University, New York University, Ohio State University, Pennsylvania State University, University of Portsmouth, Princeton University, the Spanish Participation Group, University of Tokyo, University of Utah, Vanderbilt University, University of Virginia, University of Washington, and Yale University.

The Galaxy Evolution Explorer (GALEX) is a NASA Small Explorer. The mission was developed in cooperation with the Centre National d'Etudes Spatiales of France and the Korean Ministry of Science and Technology.

This publication makes use of data products from the Wide-field Infrared Survey Explorer, which is a joint project of the University of California, Los Angeles, and the Jet Propulsion Laboratory/California Institute of Technology, funded by the National Aeronautics and Space Administration.

This research has made use of the NASA/IPAC Infrared Science Archive, which is funded by the National Aeronautics and Space Administration and operated by the California Institute of Technology.

The construction of GSWLC was funded through NASA award NNX12AE06G.

This research made use of Astropy,\footnote{http://www.astropy.org} a community-developed core Python package for Astronomy \citep{astropy13a, astropy18a}, matplotlib \citep{hunter07}, and TOPCAT \citep{taylor05a}. 

\facilities{Arecibo, IRSA, Spitzer, WISE, GALEX}
\software{IDL, astropy, matplotlib, TOPCAT}

\bibliography{mybib}{}

\begin{thebibliography}{}
\expandafter\ifx\csname natexlab\endcsname\relax\def\natexlab#1{#1}\fi
\providecommand{\url}[1]{\href{#1}{#1}}
\providecommand{\dodoi}[1]{doi:~\href{http://doi.org/#1}{\nolinkurl{#1}}}
\providecommand{\doeprint}[1]{\href{http://ascl.net/#1}{\nolinkurl{http://ascl.net/#1}}}
\providecommand{\doarXiv}[1]{\href{https://arxiv.org/abs/#1}{\nolinkurl{https://arxiv.org/abs/#1}}}

\bibitem[{{Astropy Collaboration} {et~al.}(2013){Astropy Collaboration},
  {Robitaille}, {Tollerud}, {Greenfield}, {Droettboom}, {Bray}, {Aldcroft},
  {Davis}, {Ginsburg}, {Price-Whelan}, {Kerzendorf}, {Conley}, {Crighton},
  {Barbary}, {Muna}, {Ferguson}, {Grollier}, {Parikh}, {Nair}, {Unther},
  {Deil}, {Woillez}, {Conseil}, {Kramer}, {Turner}, {Singer}, {Fox}, {Weaver},
  {Zabalza}, {Edwards}, {Azalee Bostroem}, {Burke}, {Casey}, {Crawford},
  {Dencheva}, {Ely}, {Jenness}, {Labrie}, {Lim}, {Pierfederici}, {Pontzen},
  {Ptak}, {Refsdal}, {Servillat}, \& {Streicher}}]{astropy13a}
{Astropy Collaboration}, {Robitaille}, T.~P., {Tollerud}, E.~J., {et~al.} 2013,
  \aap, 558, A33, \dodoi{10.1051/0004-6361/201322068}

\bibitem[{{Astropy Collaboration} {et~al.}(2018){Astropy Collaboration},
  {Price-Whelan}, {Sip{\H{o}}cz}, {G{\"u}nther}, {Lim}, {Crawford}, {Conseil},
  {Shupe}, {Craig}, {Dencheva}, {Ginsburg}, {Vand erPlas}, {Bradley},
  {P{\'e}rez-Su{\'a}rez}, {de Val-Borro}, {Aldcroft}, {Cruz}, {Robitaille},
  {Tollerud}, {Ardelean}, {Babej}, {Bach}, {Bachetti}, {Bakanov}, {Bamford},
  {Barentsen}, {Barmby}, {Baumbach}, {Berry}, {Biscani}, {Boquien}, {Bostroem},
  {Bouma}, {Brammer}, {Bray}, {Breytenbach}, {Buddelmeijer}, {Burke},
  {Calderone}, {Cano Rodr{\'\i}guez}, {Cara}, {Cardoso}, {Cheedella}, {Copin},
  {Corrales}, {Crichton}, {D'Avella}, {Deil}, {Depagne}, {Dietrich}, {Donath},
  {Droettboom}, {Earl}, {Erben}, {Fabbro}, {Ferreira}, {Finethy}, {Fox},
  {Garrison}, {Gibbons}, {Goldstein}, {Gommers}, {Greco}, {Greenfield},
  {Groener}, {Grollier}, {Hagen}, {Hirst}, {Homeier}, {Horton}, {Hosseinzadeh},
  {Hu}, {Hunkeler}, {Ivezi{\'c}}, {Jain}, {Jenness}, {Kanarek}, {Kendrew},
  {Kern}, {Kerzendorf}, {Khvalko}, {King}, {Kirkby}, {Kulkarni}, {Kumar},
  {Lee}, {Lenz}, {Littlefair}, {Ma}, {Macleod}, {Mastropietro}, {McCully},
  {Montagnac}, {Morris}, {Mueller}, {Mumford}, {Muna}, {Murphy}, {Nelson},
  {Nguyen}, {Ninan}, {N{\"o}the}, {Ogaz}, {Oh}, {Parejko}, {Parley}, {Pascual},
  {Patil}, {Patil}, {Plunkett}, {Prochaska}, {Rastogi}, {Reddy Janga},
  {Sabater}, {Sakurikar}, {Seifert}, {Sherbert}, {Sherwood-Taylor}, {Shih},
  {Sick}, {Silbiger}, {Singanamalla}, {Singer}, {Sladen}, {Sooley},
  {Sornarajah}, {Streicher}, {Teuben}, {Thomas}, {Tremblay}, {Turner},
  {Terr{\'o}n}, {van Kerkwijk}, {de la Vega}, {Watkins}, {Weaver}, {Whitmore},
  {Woillez}, {Zabalza}, \& {Astropy Contributors}}]{astropy18a}
{Astropy Collaboration}, {Price-Whelan}, A.~M., {Sip{\H{o}}cz}, B.~M., {et~al.}
  2018, \aj, 156, 123, \dodoi{10.3847/1538-3881/aabc4f}

\bibitem[{{Balogh} {et~al.}(2004){Balogh}, {Baldry}, {Nichol}, {Miller},
  {Bower}, \& {Glazebrook}}]{balogh04a}
{Balogh}, M.~L., {Baldry}, I.~K., {Nichol}, R., {et~al.} 2004, \apjl, 615,
  L101, \dodoi{10.1086/426079}

\bibitem[{{Blanton} {et~al.}(2005){Blanton}, {Eisenstein}, {Hogg}, {Schlegel},
  \& {Brinkmann}}]{blanton05a}
{Blanton}, M.~R., {Eisenstein}, D., {Hogg}, D.~W., {Schlegel}, D.~J., \&
  {Brinkmann}, J. 2005, \apj, 629, 143, \dodoi{10.1086/422897}

\bibitem[{{Blanton} {et~al.}(2011){Blanton}, {Kazin}, {Muna}, {Weaver}, \&
  {Price-Whelan}}]{blanton11a}
{Blanton}, M.~R., {Kazin}, E., {Muna}, D., {Weaver}, B.~A., \& {Price-Whelan},
  A. 2011, \aj, 142, 31, \dodoi{10.1088/0004-6256/142/1/31}

\bibitem[{{Blanton} \& {Roweis}(2007)}]{blanton07a}
{Blanton}, M.~R., \& {Roweis}, S. 2007, \aj, 133, 734, \dodoi{10.1086/510127}

\bibitem[{{Brinchmann} {et~al.}(2004){Brinchmann}, {Charlot}, {White},
  {Tremonti}, {Kauffmann}, {Heckman}, \& {Brinkmann}}]{brinchmann04a}
{Brinchmann}, J., {Charlot}, S., {White}, S.~D.~M., {et~al.} 2004, \mnras, 351,
  1151, \dodoi{10.1111/j.1365-2966.2004.07881.x}

\bibitem[{{Brough} {et~al.}(2013){Brough}, {Croom}, {Sharp}, {Hopkins},
  {Taylor}, {Baldry}, {Gunawardhana}, {Liske}, {Norberg}, {Robotham}, {Bauer},
  {Bland-Hawthorn}, {Colless}, {Foster}, {Kelvin}, {Lara-Lopez},
  {L{\'o}pez-S{\'a}nchez}, {Loveday}, {Owers}, {Pimbblet}, \&
  {Prescott}}]{brough13a}
{Brough}, S., {Croom}, S., {Sharp}, R., {et~al.} 2013, \mnras, 435, 2903,
  \dodoi{10.1093/mnras/stt1489}

\bibitem[{{Catinella} {et~al.}(2018){Catinella}, {Saintonge}, {Janowiecki},
  {Cortese}, {Dav{\'e}}, {Lemonias}, {Cooper}, {Schiminovich}, {Hummels},
  {Fabello}, {Ger{\'e}b}, {Kilborn}, \& {Wang}}]{Catinella18a}
{Catinella}, B., {Saintonge}, A., {Janowiecki}, S., {et~al.} 2018, \mnras, 476,
  875, \dodoi{10.1093/mnras/sty089}

\bibitem[{{Chen} {et~al.}(2012){Chen}, {Kauffmann}, {Tremonti}, {White},
  {Heckman}, {Kova{\v{c}}}, {Bundy}, {Chisholm}, {Maraston}, {Schneider},
  {Bolton}, {Weaver}, \& {Brinkmann}}]{chen12a}
{Chen}, Y.-M., {Kauffmann}, G., {Tremonti}, C.~A., {et~al.} 2012, \mnras, 421,
  314, \dodoi{10.1111/j.1365-2966.2011.20306.x}

\bibitem[{{Cluver} {et~al.}(2014){Cluver}, {Jarrett}, {Hopkins}, {Driver},
  {Liske}, {Gunawardhana}, {Taylor}, {Robotham}, {Alpaslan}, {Baldry}, {Brown},
  {Peacock}, {Popescu}, {Tuffs}, {Bauer}, {Bland -Hawthorn}, {Colless},
  {Holwerda}, {Lara-L{\'o}pez}, {Leschinski}, {L{\'o}pez-S{\'a}nchez},
  {Norberg}, {Owers}, {Wang}, \& {Wilkins}}]{cluver14}
{Cluver}, M.~E., {Jarrett}, T.~H., {Hopkins}, A.~M., {et~al.} 2014, \apj, 782,
  90, \dodoi{10.1088/0004-637X/782/2/90}

\bibitem[{{Condon} {et~al.}(1998){Condon}, {Cotton}, E.W., {Yin}, {Perley},
  {Taylor}, \& {Broderick}}]{condon98}
{Condon}, J., {Cotton}, W., E.W., G., {et~al.} 1998, \aj, 115, 1693,
  \dodoi{10.1086/300337}

\bibitem[{{Devour} \& {Bell}(2016)}]{devour16a}
{Devour}, B.~M., \& {Bell}, E.~F. 2016, \mnras, 459, 2054,
  \dodoi{10.1093/mnras/stw754}

\bibitem[{{Fossati} {et~al.}(2013){Fossati}, {Gavazzi}, {Savorgnan},
  {Fumagalli}, {Boselli}, {Guti{\'e}rrez}, {Hern{\'a}ndez Toledo},
  {Giovanelli}, \& {Haynes}}]{fossati13a}
{Fossati}, M., {Gavazzi}, G., {Savorgnan}, G., {et~al.} 2013, \aap, 553, A91,
  \dodoi{10.1051/0004-6361/201220915}

\bibitem[{{Gavazzi} {et~al.}(2013){Gavazzi}, {Fumagalli}, {Fossati}, {Galardo},
  {Grossetti}, {Boselli}, {Giovanelli}, \& {Haynes}}]{gavazzi13a}
{Gavazzi}, G., {Fumagalli}, M., {Fossati}, M., {et~al.} 2013, \aap, 553, A89,
  \dodoi{10.1051/0004-6361/201218789}

\bibitem[{{Giovanelli} {et~al.}(1994){Giovanelli}, {Haynes}, {Salzer},
  {Wegner}, {da Costa}, \& {Freudling}}]{giovanelli94a}
{Giovanelli}, R., {Haynes}, M.~P., {Salzer}, J.~J., {et~al.} 1994, \aj, 107,
  2036, \dodoi{10.1086/117014}

\bibitem[{{Giovanelli} {et~al.}(2005){Giovanelli}, {Haynes}, {Kent},
  {Perillat}, {Saintonge}, {Brosch}, {Catinella}, {Hoffman}, {Stierwalt},
  {Spekkens}, {Lerner}, {Masters}, {Momjian}, {Rosenberg}, {Springob},
  {Boselli}, {Charmandaris}, {Darling}, {Davies}, {Garcia Lambas}, {Gavazzi},
  {Giovanardi}, {Hardy}, {Hunt}, {Iovino}, {Karachentsev}, {Karachentseva},
  {Koopmann}, {Marinoni}, {Minchin}, {Muller}, {Putman}, {Pantoja}, {Salzer},
  {Scodeggio}, {Skillman}, {Solanes}, {Valotto}, {van Driel}, \& {van
  Zee}}]{giovanelli05a}
{Giovanelli}, R., {Haynes}, M.~P., {Kent}, B.~R., {et~al.} 2005, \aj, 130,
  2598, \dodoi{10.1086/497431}

\bibitem[{{Giovanelli} {et~al.}(2007){Giovanelli}, {Haynes}, {Kent},
  {Saintonge}, {Stierwalt}, {Altaf}, {Balonek}, {Brosch}, {Brown}, {Catinella},
  {Furniss}, {Goldstein}, {Hoffman}, {Koopmann}, {Kornreich}, {Mahmood},
  {Martin}, {Masters}, {Mitschang}, {Momjian}, {Nair}, {Rosenberg}, \&
  {Walsh}}]{giovanelli07a}
---. 2007, \aj, 133, 2569, \dodoi{10.1086/516635}

\bibitem[{{Green} {et~al.}(2019){Green}, {Schlafly}, {Zucker}, {Speagle}, \&
  {Finkbeiner}}]{green19a}
{Green}, G.~M., {Schlafly}, E.~F., {Zucker}, C., {Speagle}, J.~S., \&
  {Finkbeiner}, D.~P. 2019, arXiv e-prints, arXiv:1905.02734.
\newblock \doarXiv{1905.02734}

\bibitem[{{Green} {et~al.}(2015){Green}, {Schlafly}, {Finkbeiner}, {Rix},
  {Martin}, {Burgett}, {Draper}, {Flewelling}, {Hodapp}, {Kaiser}, {Kudritzki},
  {Magnier}, {Metcalfe}, {Price}, {Tonry}, \& {Wainscoat}}]{green15a}
{Green}, G.~M., {Schlafly}, E.~F., {Finkbeiner}, D.~P., {et~al.} 2015, \apj,
  810, 25, \dodoi{10.1088/0004-637X/810/1/25}

\bibitem[{{Hao} {et~al.}(2011){Hao}, {Kennicutt}, {Johnson}, {Calzetti},
  {Dale}, \& {Moustakas}}]{hao11a}
{Hao}, C.-N., {Kennicutt}, R.~C., {Johnson}, B.~D., {et~al.} 2011, \apj, 741,
  124, \dodoi{10.1088/0004-637X/741/2/124}

\bibitem[{{Haynes} {et~al.}(2011){Haynes}, {Giovanelli}, {Martin}, {Hess},
  {Saintonge}, {Adams}, {Hallenbeck}, {Hoffman}, {Huang}, {Kent}, {Koopmann},
  {Papastergis}, {Stierwalt}, {Balonek}, {Craig}, {Higdon}, {Kornreich},
  {Miller}, {O'Donoghue}, {Olowin}, {Rosenberg}, {Spekkens}, {Troischt}, \&
  {Wilcots}}]{haynes11a}
{Haynes}, M.~P., {Giovanelli}, R., {Martin}, A.~M., {et~al.} 2011, \aj, 142,
  170, \dodoi{10.1088/0004-6256/142/5/170}

\bibitem[{{Haynes} {et~al.}(2018){Haynes}, {Giovanelli}, {Kent}, {Adams},
  {Balonek}, {Craig}, {Fertig}, {Finn}, {Giovanardi}, {Hallenbeck}, {Hess},
  {Hoffman}, {Huang}, {Jones}, {Koopmann}, {Kornreich}, {Leisman}, {Miller},
  {Moorman}, {O'Connor}, {O'Donoghue}, {Papastergis}, {Troischt}, {Stark}, \&
  {Xiao}}]{haynes18a}
{Haynes}, M.~P., {Giovanelli}, R., {Kent}, B.~R., {et~al.} 2018, \apj, 861, 49,
  \dodoi{10.3847/1538-4357/aac956}

\bibitem[{{Huang} {et~al.}(2012{\natexlab{a}}){Huang}, {Haynes}, {Giovanelli},
  \& {Brinchmann}}]{huang12a}
{Huang}, S., {Haynes}, M.~P., {Giovanelli}, R., \& {Brinchmann}, J.
  2012{\natexlab{a}}, \apj, 756, 113, \dodoi{10.1088/0004-637X/756/2/113}

\bibitem[{{Huang} {et~al.}(2012{\natexlab{b}}){Huang}, {Haynes}, {Giovanelli},
  {Brinchmann}, {Stierwalt}, \& {Neff}}]{huang12b}
{Huang}, S., {Haynes}, M.~P., {Giovanelli}, R., {et~al.} 2012{\natexlab{b}},
  \aj, 143, 133, \dodoi{10.1088/0004-6256/143/6/133}

\bibitem[{Hunter(2007)}]{hunter07}
Hunter, J.~D. 2007, Computing in Science \& Engineering, 9, 90,
  \dodoi{10.1109/MCSE.2007.55}

\bibitem[{{Jarrett} {et~al.}(2011){Jarrett}, {Cohen}, {Masci}, {Wright},
  {Stern}, {Benford}, {Blain}, {Carey}, {Cutri}, {Eisenhardt}, {Lonsdale},
  {Mainzer}, {Marsh}, {Padgett}, {Petty}, {Ressler}, {Skrutskie}, {Stanford},
  {Surace}, {Tsai}, {Wheelock}, \& {Yan}}]{jarrett11}
{Jarrett}, T.~H., {Cohen}, M., {Masci}, F., {et~al.} 2011, \apj, 735, 112,
  \dodoi{10.1088/0004-637X/735/2/112}

\bibitem[{{Jarrett} {et~al.}(2013){Jarrett}, {Masci}, {Tsai}, {Petty},
  {Cluver}, {Assef}, {Benford}, {Blain}, {Bridge}, {Donoso}, {Eisenhardt},
  {Koribalski}, {Lake}, {Neill}, {Seibert}, {Sheth}, {Stanford}, \&
  {Wright}}]{jarrett13}
{Jarrett}, T.~H., {Masci}, F., {Tsai}, C.~W., {et~al.} 2013, \aj, 145, 6,
  \dodoi{10.1088/0004-6256/145/1/6}

\bibitem[{{Jones} {et~al.}(2018){Jones}, {Haynes}, {Giovanelli}, \&
  {Moorman}}]{jones18b}
{Jones}, M.~G., {Haynes}, M.~P., {Giovanelli}, R., \& {Moorman}, C. 2018,
  \mnras, 477, 2, \dodoi{10.1093/mnras/sty521}

\bibitem[{{Kauffmann} {et~al.}(2003){Kauffmann}, {Heckman}, {White}, {Charlot},
  {Tremonti}, {Brinchmann}, {Bruzual}, {Peng}, {Seibert}, {Bernardi},
  {Blanton}, {Brinkmann}, {Castander}, {Cs{\'a}bai}, {Fukugita}, {Ivezic},
  {Munn}, {Nichol}, {Padmanabhan}, {Thakar}, {Weinberg}, \&
  {York}}]{kauffmann03}
{Kauffmann}, G., {Heckman}, T.~M., {White}, S. D.~M., {et~al.} 2003, \mnras,
  341, 33, \dodoi{10.1046/j.1365-8711.2003.06291.x}

\bibitem[{{Kennicutt} \& {Evans}(2012)}]{kennicutt12a}
{Kennicutt}, R.~C., \& {Evans}, N.~J. 2012, \araa, 50, 531,
  \dodoi{10.1146/annurev-astro-081811-125610}

\bibitem[{{Kent} {et~al.}(2008){Kent}, {Giovanelli}, {Haynes}, {Martin},
  {Saintonge}, {Stierwalt}, {Balonek}, {Brosch}, \& {Koopmann}}]{kent08a}
{Kent}, B.~R., {Giovanelli}, R., {Haynes}, M.~P., {et~al.} 2008, \aj, 136, 713,
  \dodoi{10.1088/0004-6256/136/2/713}

\bibitem[{{Kourkchi} {et~al.}(2019){Kourkchi}, {Tully}, {Neill}, {Seibert},
  {Courtois}, \& {Dupuy}}]{kourkchi19a}
{Kourkchi}, E., {Tully}, R.~B., {Neill}, J.~D., {et~al.} 2019, arXiv e-prints,
  arXiv:1909.01572.
\newblock \doarXiv{1909.01572}

\bibitem[{{Lang}(2014)}]{lang14}
{Lang}, D. 2014, \aj, 147, 108, \dodoi{10.1088/0004-6256/147/5/108}

\bibitem[{{Lang} {et~al.}(2016){Lang}, {Hogg}, \& {Schlegel}}]{lang16}
{Lang}, D., {Hogg}, D.~W., \& {Schlegel}, D.~J. 2016, \aj, 151, 36,
  \dodoi{10.3847/0004-6256/151/2/36}

\bibitem[{{Leisman} {et~al.}(2017){Leisman}, {Haynes}, {Janowiecki},
  {Hallenbeck}, {J{\'o}zsa}, {Giovanelli}, {Adams}, {Bernal Neira}, {Cannon},
  {Janesh}, {Rhode}, \& {Salzer}}]{leisman17a}
{Leisman}, L., {Haynes}, M.~P., {Janowiecki}, S., {et~al.} 2017, \apj, 842,
  133, \dodoi{10.3847/1538-4357/aa7575}

\bibitem[{{Leroy} {et~al.}(2019){Leroy}, {Sandstrom}, {Lang}, {Lewis}, {Salim},
  {Behrens}, {Chastenet}, {Chiang}, {Gallagher}, {Kessler}, \&
  {Utomo}}]{leroy19a}
{Leroy}, A.~K., {Sandstrom}, K.~M., {Lang}, D., {et~al.} 2019, \apjs, 244, 24,
  \dodoi{10.3847/1538-4365/ab3925}

\bibitem[{{Maraston} {et~al.}(2013){Maraston}, {Pforr}, {Henriques}, {Thomas},
  {Wake}, {Brownstein}, {Capozzi}, {Tinker}, {Bundy}, {Skibba}, {Beifiori},
  {Nichol}, {Edmondson}, {Schneider}, {Chen}, {Masters}, {Steele}, {Bolton},
  {York}, {Weaver}, {Higgs}, {Bizyaev}, {Brewington}, {Malanushenko},
  {Malanushenko}, {Snedden}, {Oravetz}, {Pan}, {Shelden}, \&
  {Simmons}}]{maraston13a}
{Maraston}, C., {Pforr}, J., {Henriques}, B.~M., {et~al.} 2013, \mnras, 435,
  2764, \dodoi{10.1093/mnras/stt1424}

\bibitem[{{Martin} {et~al.}(2012){Martin}, {Giovanelli}, {Haynes}, \&
  {Guzzo}}]{martin12a}
{Martin}, A.~M., {Giovanelli}, R., {Haynes}, M.~P., \& {Guzzo}, L. 2012, \apj,
  750, 38, \dodoi{10.1088/0004-637X/750/1/38}

\bibitem[{{Masters} {et~al.}(2010){Masters}, {Nichol}, {Bamford}, {Mosleh},
  {Lintott}, {Andreescu}, {Edmondson}, {Keel}, {Murray}, {Raddick},
  {Schawinski}, {Slosar}, {Szalay}, {Thomas}, \& {Vandenberg}}]{masters10a}
{Masters}, K.~L., {Nichol}, R., {Bamford}, S., {et~al.} 2010, \mnras, 404, 792,
  \dodoi{10.1111/j.1365-2966.2010.16335.x}

\bibitem[{{McGaugh} \& {Schombert}(2015)}]{mcgaugh15a}
{McGaugh}, S.~S., \& {Schombert}, J.~M. 2015, The Astrophysical Journal, 802,
  18, \dodoi{10.1088/0004-637X/802/1/18}

\bibitem[{{McGaugh} {et~al.}(2000){McGaugh}, {Schombert}, {Bothun}, \& {de
  Blok}}]{mcgaugh00a}
{McGaugh}, S.~S., {Schombert}, J.~M., {Bothun}, G.~D., \& {de Blok}, W.~J.~G.
  2000, \apjl, 533, L99, \dodoi{10.1086/312628}

\bibitem[{{Moorman} {et~al.}(2014){Moorman}, {Vogeley}, {Hoyle}, {Pan},
  {Haynes}, \& {Giovanelli}}]{moorman14a}
{Moorman}, C.~M., {Vogeley}, M.~S., {Hoyle}, F., {et~al.} 2014, \mnras, 444,
  3559, \dodoi{10.1093/mnras/stu1674}

\bibitem[{{Mouhcine} {et~al.}(2007){Mouhcine}, {Baldry}, \&
  {Bamford}}]{mouhcine07a}
{Mouhcine}, M., {Baldry}, I.~K., \& {Bamford}, S.~P. 2007, \mnras, 382, 801,
  \dodoi{10.1111/j.1365-2966.2007.12405.x}

\bibitem[{{Mu{\~n}oz-Mateos} {et~al.}(2013){Mu{\~n}oz-Mateos}, {Sheth}, {Gil de
  Paz}, {Meidt}, {Athanassoula}, {Bosma}, {Comer{\'o}n}, {Elmegreen},
  {Elmegreen}, {Erroz-Ferrer}, {Gadotti}, {Hinz}, {Ho}, {Holwerda}, {Jarrett},
  {Kim}, {Knapen}, {Laine}, {Laurikainen}, {Madore}, {Menendez-Delmestre},
  {Mizusawa}, {Regan}, {Salo}, {Schinnerer}, {Seibert}, {Skibba}, \&
  {Zaritsky}}]{munos2013a}
{Mu{\~n}oz-Mateos}, J.~C., {Sheth}, K., {Gil de Paz}, A., {et~al.} 2013, \apj,
  771, 59, \dodoi{10.1088/0004-637X/771/1/59}

\bibitem[{{Murphy} {et~al.}(2011){Murphy}, {Condon}, {Schinnerer}, {Kennicutt},
  {Calzetti}, {Armus}, {Helou}, {Turner}, {Aniano}, {Beir{\~a}o}, {Bolatto},
  {Brandl}, {Croxall}, {Dale}, {Donovan Meyer}, {Draine}, {Engelbracht},
  {Hunt}, {Hao}, {Koda}, {Roussel}, {Skibba}, \& {Smith}}]{murphy11a}
{Murphy}, E.~J., {Condon}, J.~J., {Schinnerer}, E., {et~al.} 2011, \apj, 737,
  67, \dodoi{10.1088/0004-637X/737/2/67}

\bibitem[{{Papastergis} {et~al.}(2016){Papastergis}, {Adams}, \& {van der
  Hulst}}]{papastergis16a}
{Papastergis}, E., {Adams}, E.~A.~K., \& {van der Hulst}, J.~M. 2016, \aap,
  593, A39, \dodoi{10.1051/0004-6361/201628410}

\bibitem[{{Papastergis} {et~al.}(2013){Papastergis}, {Giovanelli}, {Haynes},
  {Rodr{\'{\i}}guez-Puebla}, \& {Jones}}]{papastergis13a}
{Papastergis}, E., {Giovanelli}, R., {Haynes}, M.~P.,
  {Rodr{\'{\i}}guez-Puebla}, A., \& {Jones}, M.~G. 2013, \apj, 776, 43,
  \dodoi{10.1088/0004-637X/776/1/43}

\bibitem[{{Papastergis} {et~al.}(2011){Papastergis}, {Martin}, {Giovanelli}, \&
  {Haynes}}]{papastergis11a}
{Papastergis}, E., {Martin}, A.~M., {Giovanelli}, R., \& {Haynes}, M.~P. 2011,
  \apj, 739, 38, \dodoi{10.1088/0004-637X/739/1/38}

\bibitem[{{Planck Collaboration} {et~al.}(2014){Planck Collaboration},
  {Abergel}, {Ade}, {Aghanim}, {Alves}, {Aniano}, {Armitage-Caplan}, {Arnaud},
  {Ashdown}, {Atrio-Barand ela}, \& et~al.}]{planck14a}
{Planck Collaboration}, {Abergel}, A., {Ade}, P.~A.~R., {et~al.} 2014, \aap,
  571, A11, \dodoi{10.1051/0004-6361/201323195}

\bibitem[{{Querejeta} {et~al.}(2015){Querejeta}, {Meidt}, {Schinnerer},
  {Cisternas}, {Mu{\~n}oz-Mateos}, {Sheth}, {Knapen}, {van de Ven}, {Norris},
  {Peletier}, {Laurikainen}, {Salo}, {Holwerda}, {Athanassoula}, {Bosma},
  {Groves}, {Ho}, {Gadotti}, {Zaritsky}, {Regan}, {Hinz}, {Gil de Paz},
  {Menendez-Delmestre}, {Seibert}, {Mizusawa}, {Kim}, {Erroz-Ferrer}, {Laine},
  \& {Comer{\'o}n}}]{querejeta15a}
{Querejeta}, M., {Meidt}, S.~E., {Schinnerer}, E., {et~al.} 2015, \apjs, 219,
  5, \dodoi{10.1088/0067-0049/219/1/5}

\bibitem[{{Rieke} {et~al.}(2009){Rieke}, {Alonso-Herrero}, {Weiner},
  {P{\'e}rez-Gonz{\'a}lez}, {Blaylock}, {Donley}, \& {Marcillac}}]{rieke09a}
{Rieke}, G.~H., {Alonso-Herrero}, A., {Weiner}, B.~J., {et~al.} 2009, \apj,
  692, 556, \dodoi{10.1088/0004-637X/692/1/556}

\bibitem[{{Saintonge}(2007)}]{saintonge07a}
{Saintonge}, A. 2007, \aj, 133, 2087, \dodoi{10.1086/513515}

\bibitem[{{Saintonge} {et~al.}(2017){Saintonge}, {Catinella}, {Tacconi},
  {Kauffmann}, {Genzel}, {Cortese}, {Dav{\'e}}, {Fletcher},
  {Graci{\'a}-Carpio}, {Kramer}, {Heckman}, {Janowiecki}, {Lutz}, {Rosario},
  {Schiminovich}, {Schuster}, {Wang}, {Wuyts}, {Borthakur}, {Lamperti}, \&
  {Roberts-Borsani}}]{Saintonge17}
{Saintonge}, A., {Catinella}, B., {Tacconi}, L.~J., {et~al.} 2017, \apjs, 233,
  22, \dodoi{10.3847/1538-4365/aa97e0}

\bibitem[{{Salim} {et~al.}(2018){Salim}, {Boquien}, \& {Lee}}]{salim18a}
{Salim}, S., {Boquien}, M., \& {Lee}, J.~C. 2018, \apj, 859, 11,
  \dodoi{10.3847/1538-4357/aabf3c}

\bibitem[{{Salim} {et~al.}(2016){Salim}, {Lee}, {Janowiecki}, {da Cunha},
  {Dickinson}, {Boquien}, {Burgarella}, {Salzer}, \& {Charlot}}]{salim16a}
{Salim}, S., {Lee}, J.~C., {Janowiecki}, S., {et~al.} 2016, \apjs, 227, 2,
  \dodoi{10.3847/0067-0049/227/1/2}

\bibitem[{{Schlafly} \& {Finkbeiner}(2011)}]{schlafly11a}
{Schlafly}, E.~F., \& {Finkbeiner}, D.~P. 2011, \apj, 737, 103,
  \dodoi{10.1088/0004-637X/737/2/103}

\bibitem[{{Schlafly} {et~al.}(2019){Schlafly}, {Meisner}, \&
  {Green}}]{schlafly19}
{Schlafly}, E.~F., {Meisner}, A.~M., \& {Green}, G.~M. 2019, \apjs, 240, 30,
  \dodoi{10.3847/1538-4365/aafbea}

\bibitem[{{Schlegel} {et~al.}(1998){Schlegel}, {Finkbeiner}, \&
  {Davis}}]{schlegel1998a}
{Schlegel}, D.~J., {Finkbeiner}, D.~P., \& {Davis}, M. 1998, \apj, 500, 525,
  \dodoi{10.1086/305772}

\bibitem[{{Shao} {et~al.}(2007){Shao}, {Xiao}, {Shen}, {Mo}, {Xia}, \&
  {Deng}}]{shao07}
{Shao}, Z., {Xiao}, Q., {Shen}, S., {et~al.} 2007, \apj, 659, 1159,
  \dodoi{10.1086/511131}

\bibitem[{{Sheth} {et~al.}(2010){Sheth}, {Regan}, {Hinz}, {Gil de Paz},
  {Men{\'e}ndez-Delmestre}, {Mu{\~n}oz-Mateos}, {Seibert}, {Kim},
  {Laurikainen}, {Salo}, {Gadotti}, {Laine}, {Mizusawa}, {Armus},
  {Athanassoula}, {Bosma}, {Buta}, {Capak}, {Jarrett}, {Elmegreen},
  {Elmegreen}, {Knapen}, {Koda}, {Helou}, {Ho}, {Madore}, {Masters},
  {Mobasher}, {Ogle}, {Peng}, {Schinnerer}, {Surace}, {Zaritsky},
  {Comer{\'o}n}, {de Swardt}, {Meidt}, {Kasliwal}, \& {Aravena}}]{sheth10}
{Sheth}, K., {Regan}, M., {Hinz}, J.~L., {et~al.} 2010, \pasp, 122, 1397,
  \dodoi{10.1086/657638}

\bibitem[{{Taylor} {et~al.}(2011){Taylor}, {Hopkins}, {Baldry}, {Brown},
  {Driver}, {Kelvin}, {Hill}, {Robotham}, {Bland-Hawthorn}, {Jones}, {Sharp},
  {Thomas}, {Liske}, {Loveday}, {Norberg}, {Peacock}, {Bamford}, {Brough},
  {Colless}, {Cameron}, {Conselice}, {Croom}, {Frenk}, {Gunawardhana},
  {Kuijken}, {Nichol}, {Parkinson}, {Phillipps}, {Pimbblet}, {Popescu},
  {Prescott}, {Sutherland}, {Tuffs}, {van Kampen}, \& {Wijesinghe}}]{taylor11a}
{Taylor}, E.~N., {Hopkins}, A.~M., {Baldry}, I.~K., {et~al.} 2011, \mnras, 418,
  1587, \dodoi{10.1111/j.1365-2966.2011.19536.x}

\bibitem[{{Taylor}(2005)}]{taylor05a}
{Taylor}, M.~B. 2005, in Astronomical Society of the Pacific Conference Series,
  Vol. 347, Astronomical Data Analysis Software and Systems XIV, ed.
  P.~{Shopbell}, M.~{Britton}, \& R.~{Ebert}, 29

\bibitem[{{Tully} \& {Fisher}(1977)}]{tully77a}
{Tully}, R.~B., \& {Fisher}, J.~R. 1977, Astronomy and Astrophysics, 500, 105

\bibitem[{{Tully} {et~al.}(1998){Tully}, {Pierce}, {Huang}, {Saunders},
  {Verheijen}, \& {Witchalls}}]{tully98a}
{Tully}, R.~B., {Pierce}, M.~J., {Huang}, J.-S., {et~al.} 1998, \aj, 115, 2264,
  \dodoi{10.1086/300379}

\bibitem[{{Wright} {et~al.}(2010){Wright}, {Eisenhardt}, {Mainzer}, {Ressler},
  {Cutri}, {Jarrett}, {Kirkpatrick}, {Padgett}, {McMillan}, {Skrutskie},
  {Stanford}, {Cohen}, {Walker}, {Mather}, {Leisawitz}, {Gautier}, {McLean},
  {Benford}, {Lonsdale}, {Blain}, {Mendez}, {Irace}, {Duval}, {Liu}, {Royer},
  {Heinrichsen}, {Howard}, {Shannon}, {Kendall}, {Walsh}, {Larsen}, {Cardon},
  {Schick}, {Schwalm}, {Abid}, {Fabinsky}, {Naes}, \& {Tsai}}]{wright10}
{Wright}, E.~L., {Eisenhardt}, P. R.~M., {Mainzer}, A.~K., {et~al.} 2010, \aj,
  140, 1868, \dodoi{10.1088/0004-6256/140/6/1868}

\end{thebibliography}
\bibliographystyle{aasjournal}


\end{document}